\documentclass[iop,apj,numberedappendix]{emulateapj}
\pdfoutput=1
\usepackage{color}
\usepackage[version=3]{mhchem}
\usepackage{amsmath}
\usepackage{bm}
\usepackage{upgreek}
\usepackage{comment}
\usepackage{booktabs}
\usepackage{tabularx}
\usepackage{natbib}
\usepackage{hyperref}
\usepackage[capitalize]{cleveref}

\newcommand{\newacronym}[3]{\newcommand{#1}[1]{#3##1 (#2##1)%
\renewcommand{#1}[1]{#2####1}}}

\newacronym{\NSE}{NSE}{nuclear statistical equilibrium}
\newacronym{\EOS}{EOS}{equation of state}
\newacronym{\sGRB}{sGRB}{short gamma ray burst}
\newacronym{\HST}{HST}{Hubble Space Telescope}

\newcommand{\kb}{\smash{k_B\,\text{baryon}^{-1}}}
\newcommand{\web}{\url{http://stellarcollapse.org/lippunerroberts2015}}
\newcommand{\skynet}{\emph{SkyNet}}
\newcommand{\smce}[1]{\smash{\ce{#1}}}

\crefformat{footnote}{#2\footnotemark[#1]#3}
\crefname{equation}{Equation}{Equations}
\crefname{figure}{Figure}{Figures}

\newcommand\sref[1]{Section~\ref{#1}}

\slugcomment{Final version of November 2, 2015}

\hypersetup{
  pdfauthor={Jonas Lippuner and Luke Roberts},
  pdftitle={r-Process Lanthanide Production and Heating Rates in Kilonovae},
  pdfkeywords={gamma-ray burst: general; gravitational waves;
nuclear reactions, nucleosynthesis, abundances; stars: neutron}
}

\begin{document}

\title{r-Process Lanthanide Production and Heating Rates in Kilonovae}
\shorttitle{Lanthanides and Heating Rates in Kilonovae}

\shortauthors{Lippuner and Roberts}

\author{Jonas Lippuner and Luke F. Roberts\altaffilmark{1}}
\affil{TAPIR, Walter Burke Institute for Theoretical Physics, California
Institute of Technology, MC 350-17,
1200 E California Blvd, Pasadena CA 91125, USA}

\email{jlippuner@tapir.caltech.edu}

\altaffiltext{1}{NASA Einstein Fellow}

\begin{abstract}

r-Process nucleosynthesis in material ejected during neutron star mergers may
lead to radioactively powered transients called kilonovae. The timescale and
peak luminosity of these transients depend on the composition of the
ejecta, which determines the local heating rate from
nuclear decays and the opacity. Kasen et al.\ (2013, ApJ, 774, 25) and Tanaka \&
Hotokezaka (2013, ApJ, 775, 113) pointed out that lanthanides can drastically
increase the opacity in these outflows. We use the new general-purpose nuclear
reaction network \skynet\ to carry out a parameter study of r-process
nucleosynthesis for a range of initial electron fractions $Y_e$, initial
specific entropies $s$, and expansion timescales $\tau$. We find that the ejecta
is lanthanide-free for $Y_e \gtrsim 0.22 - 0.30$, depending on $s$ and $\tau$.
The heating rate is insensitive to $s$ and $\tau$, but certain, larger values of
$Y_e$ lead to reduced heating rates, due to individual nuclides dominating the
heating. We calculate approximate light curves with a simplified gray
radiative transport scheme. The light curves peak at
about a day (week) in the lanthanide-free (-rich) cases. The heating rate does
not change much as the ejecta
becomes lanthanide-free with increasing $Y_e$, but the light curve peak becomes
about an order of magnitude brighter because it peaks much earlier when the
heating rate is larger. We also provide parametric fits for the heating rates
between 0.1 and $100\,\text{days}$, and we provide a simple fit in $Y_e$, $s$,
and $\tau$ to estimate whether the ejecta is lanthanide-rich or not.

\end{abstract}


\section{Introduction}

%
%

The merger of a compact binary system that includes at least one neutron star,
hence the merger of a neutron star with a black hole (NSBH) or the merger of two
neutron stars (NSNS), is likely to eject a significant amount of
material during the final stages of coalescence
\citep{lattimer:77} in addition to emitting gravitational waves that may be
observed by gravitational wave detectors such as advanced LIGO \citep{aligo}
and possibly
powering \sGRB{s} \citep[e.g.][]{lee_rev:07,nakar:07a,gehrels:09}. The
material that is unbound during the merger is of
interest for two main reasons.  First, the majority of the mass ejected in these
events is very neutron-rich.  Once the material decompresses from initial
densities close to nuclear density, the large number of neutrons can rapidly
capture on the few heavy nuclides present and produce nuclei up to nuclear
mass 300. This process is called the r-process because neutrons are captured
rapidly compared to the $\beta$-decay timescale of the unstable nuclides
produced by neutron capture. Thus the r-process quickly creates heavy, very
neutron-rich nuclides that eventually decay back to stability after the neutron
capture ceases \citep{burbidge:57}. Depending on the rate of NSBH and NSNS
mergers
and the
amount of neutron-rich material ejected during these events, they can be
the dominant source of r-process nucleosynthesis in the universe
\citep{argast:04,shen:14,vandevoort:15,ramirez-ruiz:15}.

Second, observable electromagnetic signals may be associated with these ejecta.
A
radio transient that occurs on a timescale of a few weeks can be powered by the
interaction of the ejecta with the surrounding medium \citep{nakar:11}.
Additionally, radioactive decay of unstable nuclides formed during decompression
of the ejecta can power a transient in the optical or infrared that peaks on a
timescale of a day to a week \citep{li:98,kulkarni:05,metzger:10,kasen:13,
tanaka:13}.  These are often referred to as either ``kilonovae''
\citep{metzger:10} or ``macronovae''\citep{kulkarni:05}.
In fact, one of these events may have been observed.  An excess in the
infrared afterglow of nearby GRB130603B, which was an \sGRB{}, has been
interpreted by some authors as a strong indicator of a transient powered by the
decay of r-process material \citep{tanvir:13,berger:13}.  A similar kilonova
like excess has recently been observed in the afterglow of GRB060614
\citep{yang:15, jin:15}.

Although almost all of the ejected material will be neutron-rich, there can be a
significant spread in the electron fraction of this neutron-rich material. The
composition will depend on whether the material was ejected tidally
\citep{lattimer:77,freiburghaus:99}, dynamically from the region where the two
neutron stars
collide \citep{bauswein:13,hotokezaka:13}, or from the accretion disk that forms
after the merger \citep{fernandez:13,perego:14,just:15}.  Since the material
ejected by all of these mechanisms starts out as cold, catalyzed material in a
neutron star, the final electron fraction
of the material depends on the weak
interaction timescale relative to the dynamical timescale of the ejecta.  If the
temperature and local neutrino density are low, and therefore weak interactions
are slow, the electron fraction is unaltered.  This is the case for the tidal
ejecta, which is predicted to have a very low electron fraction
\citep{korobkin:12}.  Conversely, material ejected from the disk stays near the
compact object for a long period and can achieve beta-equilibrium at lower
density and higher temperature \citep{just:15,richers:15}.  The dynamical
ejecta from the
contact region sits somewhere in between, as it is ejected rapidly but shocked
to
high temperatures and irradiated strongly by neutrinos, which can significantly
alter the initial electron fraction \citep{wanajo:14, goriely:15}.

At low initial electron fractions \smash{($Y_e \lesssim 0.2$)}, the final
composition of the ejecta is
relatively insensitive to the initial electron fraction of the material because
a strong r-process occurs and fission cycling produces a robust pattern
\citep{metzger:10,roberts:11,goriely:11}.  But for higher electron
fractions \smash{($0.2\lesssim Y_e \lesssim 0.3$)}, an
incomplete r-process can occur and the composition will be much more sensitive
to the properties of the outflow \citep{korobkin:12,grossman:14,kasen:15}.  In addition to
the total mass and velocity of the ejecta, the composition of the ejecta at
around a day---which determines the nuclear heating rate and opacity of the
material---plays a large role in determining the properties of the kilonova
\citep{li:98}.  Since losses due to adiabatic expansion rob all of the initial
energy from the outflow, almost all of the energy that powers the transient must
come from thermalizing the products of nuclear decay \citep{metzger:10}.  This
in turn implies that the peak luminosity of a kilonova is sensitive to the
composition.

The opacity of the material determines the timescale on which the ejecta becomes
optically thin and therefore the timescale on which the transient will peak.
\cite{kasen:13} and \cite{tanaka:13} have shown that continuum opacity is very
sensitive to the presence of lanthanides, and possibly actinides, in the
outflow.  Due to their large atomic complexity, lanthanides and actinides
have a very large number of lines relative to iron group elements and therefore
their presence drastically increases the opacity of the material and causes
predicted kilonovae to peak on timescales of around a week
\citep{barnes:13,tanaka:13}.  Older models that assumed iron-like opacities
predicted a peak timescale of around a day
\citep{metzger:10,roberts:11,goriely:11}.  Significant lanthanide and actinide
production requires very neutron-rich conditions, so \cite{metzger:14} have
suggested that measurement of the peak time of a kilonova might provide insight
into the composition of the outflow.

In this work, we present a parameter study of detailed nucleosynthesis
calculations in NSBH or NSNS merger scenarios and their associated kilonova
light curves. We focus in particular on the mass fraction of lanthanides and
actinides
present in the ejecta, the radioactive heating rate at $1\,\text{day}$,
and how these properties depend on the initial conditions of the
outflow.  As expected, the lanthanide and actinide abundances depend strongly on
the electron fraction, but the entropy and expansion timescale can also play an
important role in certain cases.  In contrast, we find that the nuclear decay
heating rate does not depend as strongly on the initial electron fraction and it
changes relatively smoothly when going from lanthanide-rich to lanthanide-free
cases.   The peak timescale, peak luminosity, and spectral temperature of our
modeled kilonovae differ substantially due to the effect of the lanthanides and
actinides on the opacity. In some cases, we also find very early and bright
transients due to a neutron-rich freeze-out, which was proposed by
\cite{kulkarni:05} and \cite{metzger:14b}.

In \sref{sec:nucleosynthesis}, we describe our parametrized nucleosynthesis
calculations and discuss how lanthanide production and the nuclear heating rate
varies over our chosen parameter space.  In \sref{sec:light_curves}, we present
simplified kilonova lightcurve models and examine how these
transients vary with outflow properties. We then conclude in
\sref{sec:conclusion}. Lanthanides and actinides both have open
\mbox{$f$-shells} and thus a similar valence electron structure, which means
their impact on the opacity is similar \citep{kasen:13}. Therefore, we will use
the term ``lanthanides'' to refer to both lanthanides and/or actinides, unless
otherwise noted.

\section{Parameterized ejecta nucleosynthesis}\label{sec:nucleosynthesis}

The details of the r-process abundance pattern, especially the position of the
third peak, can be sensitive to the nuclear mass model, reaction rates, and
fission fragment distributions that are used
\citep[e.g.][]{goriely:05,arcones:11,mumpower:12,mendoza:14,eichler:14}.  Here,
we are less interested in the detailed final abundance patterns at high mass and
more interested in the surfaces in our parameter space at which lanthanide
production ceases.  Therefore, we employ a single mass model and set of reaction
rates.  We use two models for fission fragments, but our main
results are
insensitive to this choice.

Rather than post-processing full hydrodynamic models as was done in
\citet{goriely:11,korobkin:12,grossman:14,wanajo:14,just:15,martin:15}, we use a
parametrized approach that allows us to systematically study the impact of
different ejecta properties on the properties of the ejected material relevant
to kilonovae. \citet{kasen:15} performed preliminary investigations of the
electron fraction at which lanthanide production ceases, but they did not
investigate how this influences the nuclear decay heating rate and only
considered a small region of the parameter space.

We use the following three parameters to characterize the expanding material
that undergoes r-process nucleosynthesis and produces a kilonova.

\textbf{(i)} The initial \textbf{electron fraction} $Y_e = N_p / N_B$, where
$N_p$ is the total number of protons (free or inside nuclei) and $N_B$ is the
total number of baryons. We sample $Y_e$
uniformly between 0.01 (very neutron-rich matter) and 0.5 (symmetric matter).
We do not consider $Y_e > 0.5$ because the r-process requires a neutron-rich
environment.

\textbf{(ii)} The initial \textbf{specific entropy} $s$, which we sample
logarithmically
between 1 and $100\,\kb$.

\textbf{(iii)} The \textbf{expansion timescale} $\tau$, which determines how
fast the density decreases during nuclear burning. We sample $\tau$
logarithmically between 0.1 and $500\,\text{ms}$. We choose an analytic density
profile that initially decreases exponentially with time, i.e.\
\smash{$\rho\propto
e^{-t/\tau}$}, and then transitions smoothly to a homologous,
\smash{$\rho\propto
t^{-3}$}, expansion. Requiring continuity of $\rho$ and
$d\rho/dt$ fixes the
matching point at $t=3\tau$ and gives
\begin{align}
\rho(t) = \left\{\begin{array}{ll}
\rho_0 e^{-t/\tau} & \text{if $t \leq 3\tau$,} \\
\rho_0 \left(\dfrac{3\tau}{et}\right)^3 & \text{if $t \geq 3\tau$,}
\end{array}\right. \label{eq:rho}
\end{align}
where $\rho_0$ is the initial density and $e$ is Euler's number.  This
parameterization of the density is chosen because it gives us direct control
over the dynamical timescale at the time of r-process nucleosynthesis but still
matches smoothly to the density profile expected for homologous ejecta.  We have
also found that this profile gives a good approximation to density histories of
Lagrangian fluid elements in the ejecta of BHNS mergers simulations
\citep{duez_private:15,foucart:14a}

We determine $\rho_0$ by setting the initial temperature to $T = 6\times
10^9\,\text{K}$ and then finding the density for which \NSE{} (with the given
$Y_e$)
produces a set of abundances that has the prescribed initial entropy $s$. The
entropy is calculated from the \NSE{} distribution using a modified version of
the Helmholtz \EOS{} based on \cite{timmes:00}. The \EOS{} has been modified to
calculate the entropy for each nuclear species separately, rather than using
average mass and charge numbers, and it also includes the internal partition
functions of all nuclear species, which we obtained from the WebNucleo
database distributed\footnote{\url{https://groups.nscl.msu.edu/jina/%
reaclib/db/library.php?action=viewsnapshots}} with REACLIB (see below). The
resulting initial densities
range from
$7.1\times 10^5$ to $1.4\times 10^{12}\,\text{g}\,\text{cm}^{-3}$.

Given $Y_e$, $s$, and $\tau$, \NSE{} determines $\rho_0$ (and thus $\rho(t)$)
and the initial abundances. We then use the newly developed nuclear reaction network
\skynet\ for the abundance evolution. \skynet\ is a general-purpose, modular
nuclear reaction network that keeps track of entropy and temperature changes due
to the nuclear reactions it is evolving. A detailed code description of the
functionality and features of \skynet\ is forthcoming \citep[in
prep.]{lippuner:15b}, and the source code will
be publicly released together with that paper. In the meantime, anyone who
wishes to use \skynet\ can contact the authors and request access
to the code.

We run \skynet\ with nuclear reaction rates from the JINA REACLIB
database\footnote{At the time of writing, the latest REACLIB snapshot
(2013-04-02) contains 83 incorrect $\beta$-decay rates, which we corrected for
this study. It appears that some lower limits of the half-lives published in
the Nuclear Wallet Cards (\url{http://www.nndc.bnl.gov/wallet}) were put into
REACLIB, but those lower limits can be very far away from realistic
estimates of the half-lives. For example, REACLIB gives a half-life of
$300\,\text{ns}$ for \smce{^{216}Pb} because the Nuclear Wallet Cards state the
half-life is ``$>300\,\text{ns}$'', but \citet{moller:03} gives a half-life of
about $850\,\text{s}$, which is much closer to the half-lives of similar
nuclides.} \citep{cyburt:10}. The nuclear data (masses and partition functions)
were taken from the associated WebNucleo XML file distributed with REACLIB.
Although REACLIB includes inverse rates for the strong reactions, \skynet\
calculates these inverse rates from detailed balance, so that the rates are
consistent with \NSE{}. We also include different sets of spontaneous and
neutron-induced fission rates, as REACLIB does not presently include any fission
reactions. There are three sets of symmetric neutron-induced fission reactions:
\texttt{sym0}, \texttt{sym2}, and \texttt{sym4}, which produce 0, 2, and 4 free
neutrons, respectively, for each fission event. There is also a set
\texttt{nonsym} of non-symmetric fission reactions that do not produce any free
neutrons. Each nucleosynthesis calculation includes one of the four
neutron-induced fission reaction sets and the spontaneous fission reaction set.
All the
fission reactions and their rates are taken from the same sources used in
\cite{roberts:11}.

We use beta-decay and electron capture rates from \cite{fuller:82},
\cite{Oda:94} and \cite{langanke:00} whenever they are available.  For nuclei
for which these rates are not available, the effects of electron blocking and
positron capture are approximately included by assuming that only a
ground state to ground state transition occurs as described in
\cite{arcones:10}.  These rates are then normalized such that they are
equal to the vacuum decay rates given in REACLIB at low temperature and
density, which can be thought of as setting the effective matrix element
for the ground state to ground state transition.  Because this procedure
assumes a maximal $Q$-value for these weak rates, this provides
a lower limit on the effect of the surrounding medium on the
combined beta-decay and lepton capture rate. For this study, we
run \skynet\ with 7843 nuclear species, ranging up to $Z = 112$
and $A = 337$, and 110,793 nuclear reactions.

\subsection{Parameter space} \label{sec:parameter_space}

\begin{table}
\caption{Parameter Values at Grid Points}
\label{tab:params}
\centering
\begin{tabularx}{\linewidth}{c@{\ \,}c@{\ \,}c@{}X@{}c@{\ \,}c@{\ \,}c}
\toprule
\multicolumn{3}{c}{} &&
\multicolumn{3}{c}{Additional} \\
\multicolumn{3}{c}{Low-resolution points\tablenotemark{a}} &&
\multicolumn{3}{c}{high-resolution points\tablenotemark{b}} \\
\cmidrule{1-3} \cmidrule{5-7}
$Y_e$ & $s$ & $\tau$ && $Y_e$ & $s$ & $\tau$ \\
& ($\kb$) & (ms) && & ($\kb$) & (ms)\\
\midrule
0.01 & \phn\phn1.0     & \phn\phn0.10       &&      &                 &       \\
     &                 &                    && 0.04 & \phn\phn1.3     &
\phn\phn0.17  \\
0.07 & \phn\phn1.8     & \phn\phn0.29       &&      &                 &       \\
     &                 &                    && 0.10 & \phn\phn2.4     &
\phn\phn0.49  \\
0.13 & \phn\phn3.2     & \phn\phn0.84       &&      &                 &       \\
     &                 &                    && 0.16 & \phn\phn4.2     &
\phn\phn1.4\phn   \\
0.19 & \phn\phn5.6     & \phn\phn2.4\phn    &&      &                 &       \\
     &                 &                    && 0.22 & \phn\phn7.5     &
\phn\phn4.2\phn   \\
0.25 & \phn10\phd\phn  & \phn\phn7.1\phn    &&      &                 &       \\
     &                 &                    && 0.29 & \phn13\phd\phn  &
\phn12\phd\phn\phn \\
0.32 & \phn18\phd\phn  & \phn21\phd\phn\phn &&      &                 &       \\
     &                 &                    && 0.35 & \phn24\phd\phn  &
\phn35\phd\phn\phn  \\
0.38 & \phn32\phd\phn  & \phn59\phd\phn\phn &&      &                 &       \\
     &                 &                    && 0.41 & \phn42\phd\phn  &
100\phd\phn\phn   \\
0.44 & \phn56\phd\phn  & 170\phd\phn\phn    &&      &                 &       \\
     &                 &                    && 0.47 & \phn75\phd\phn  &
290\phd\phn\phn   \\
0.50 & 100\phd\phn     & 500\phd\phn\phn    &&      &                 &       \\
 \bottomrule
\end{tabularx}
\begin{tabularx}{\linewidth}{>{\raggedright\arraybackslash}X}
${}^\text{a}$ The low-resolution runs of the entire parameter space use
only these grid points. \\
${}^\text{b}$ For the high-resolution runs of the entire parameter space we
double the number of grid points. The high-resolution runs include the
grid points shown in this column in addition to the the same points as the
low-resolution runs. \\
\end{tabularx}
\end{table}

\begin{figure*}
\includegraphics[width=\textwidth]{%
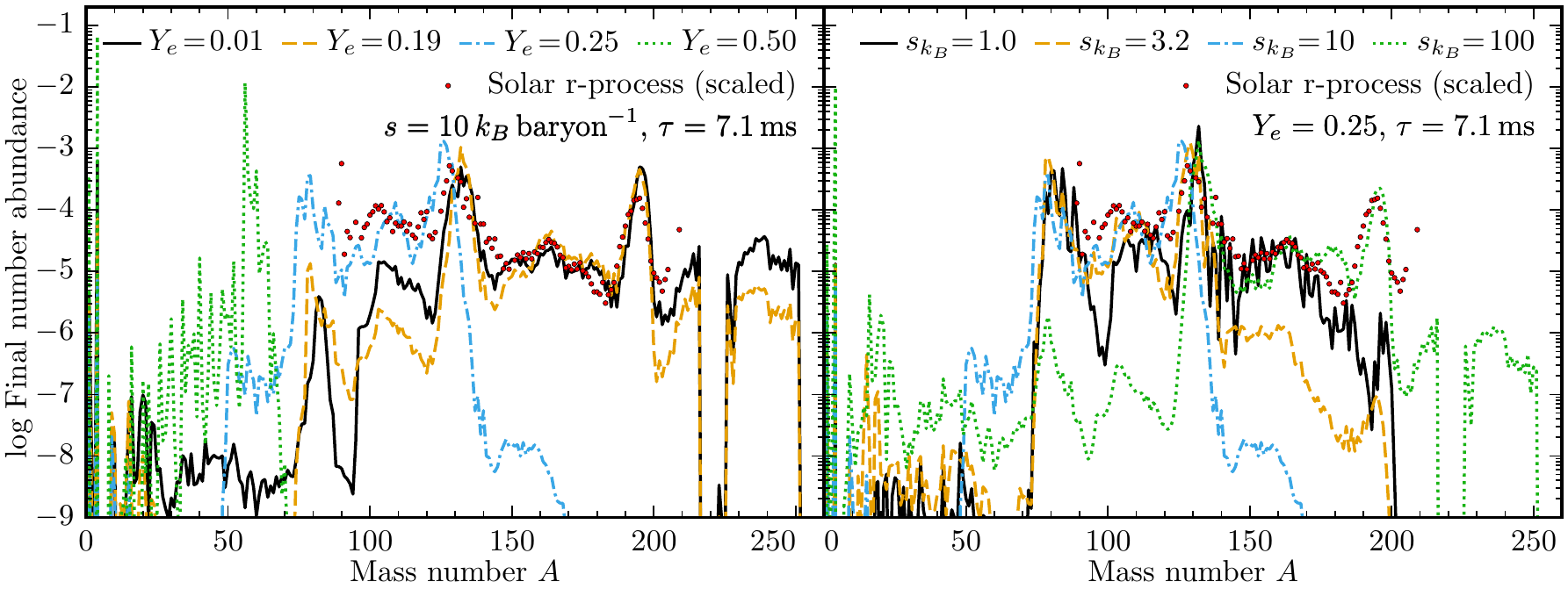}
\caption[]{The final abundances of some selected nucleosynthesis calculations.
\textbf{Left:} $Y_e = 0.01, 0.19, 0.25, 0.50$, $s = 10\,\kb$, and $\tau =
7.1\,\text{ms}$. The full r-process is made, with substantial amounts of
lanthanides and actinides, for $Y_e = 0.01$ and $Y_e = 0.19$. The $Y_e = 0.25$
trajectory is neutron-rich enough to make the second r-process peak, but not the
third and not a significant amount of lanthanides. In the symmetric case ($Y_e =
0.5$), mostly \smce{^4He} and iron-peak elements are produced. \textbf{Right:}
$Y_e = 0.25$, $s = 1.0, 3.2, 10, 100\,\kb$, and $\tau = 7.1\,\text{ms}$. With $s
= 1\,\kb$ a jagged r-process is obtained because there are only few free neutrons
per seed nucleus available and nuclides with even neutron numbers are favored.
Even though there are not many free neutrons available, there is still a
significant amount of lanthanides in the $s = 1\,\kb$ case because the initial
seed nuclei are very heavy. At higher entropies, the initial seeds become
lighter and the initial free neutron abundance increases. However, the increase
in the initial free neutron abundance is not enough to offset the decrease in
the initial mass of the seeds and so we obtain a less complete r-process. The
situation is reversed at $s = 100\,\kb$, where there is a very high
neutron-to-seed ratio. In that case, a significant fraction of $\alpha$
particles are also captured on the seed nuclei.  This leads to a full r-process
in the $s = 100\,\kb$ case.}
\label{fig:abundances} \end{figure*}

\begin{figure*}
\includegraphics[width=\textwidth]{%
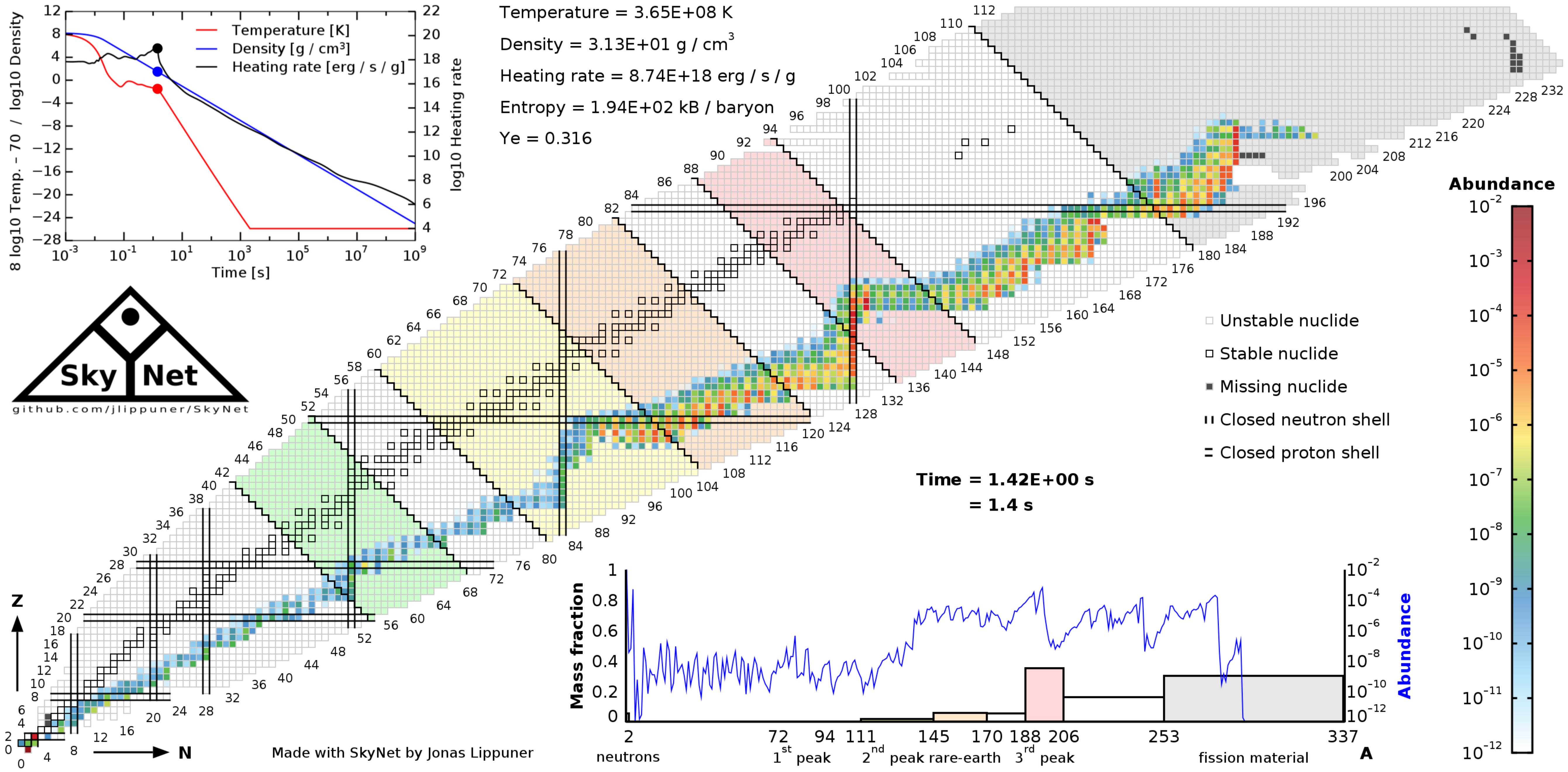}
\caption[]{A frame from the animation of the nucleosynthesis calculation for
$Y_e = 0.01$, $s = 10\,\kb$, and $\tau = 7.1\,\text{ms}$. The frame shows the
full extent of the r-process just when free neutrons get exhausted. The plot in
the upper left corner shows the temperature, density, and heating rate as
function of time. The colored bands in the chart of nuclides correspond to the
mass bins in the histogram at the bottom. The histogram shows the mass fractions
on a linear scale while the blue curve shows the abundances as a function of
mass on a logarithmic scale. The full animations are available at \web.}
\label{fig:movie_frame}
\end{figure*}

We use a $9\times9\times 9$ grid to cover the entire parameter space and run
\skynet\ for each point with all four sets of neutron-induced fission reactions
(\texttt{sym0}, \texttt{sym2}, \texttt{sym4}, \texttt{nonsym}). We also run the
\texttt{sym0} fission reactions with a finer $17\times17\times17$ grid. The
parameter values at the grid points are shown in \cref{tab:params}. The
different fission reactions only result in small quantitative and no qualitative
differences. Thus we only discuss and show plots of the high-resolution
\texttt{sym0} runs. Finally, we carry out a set of runs with high $Y_e$
resolution ($\Delta Y_e = 0.005$ resulting in 99 $Y_e$ points) for $s = 1, 10,
30, 100\,\kb$ and $\tau = 0.1, 1, 10\,\text{ms}$ with the \texttt{sym0} fission
reactions. The data underlying all the results shown and discussed here
(nucleosynthesis results, heating rate fit coefficients, light curve model
results, and integrated fractional heating contributions of all nuclides) are
available at \web.

\Cref{fig:abundances} shows the final abundances of a few selected cases, which
span the whole range of $Y_e$ and $s$ at intermediate values of the other two
parameters. For the $s = 10\,\kb$ and $\tau = 7.1\,\text{ms}$ trajectories (left
panel of \cref{fig:abundances}), the full r-process up to the third peak ($A
\sim 190$) for $Y_e = 0.01$ and $Y_e = 0.19$ is produced. We note good agreement
of the second, third, and rare-earth peak positions with the solar r-process
abundances,
although the third peak is slightly overproduced relative to the second peak.
The abundance patterns of $Y_e = 0.01$ and $Y_e = 0.19$ are very similar because
both cases are neutron-rich enough to produce nuclides with $A \gtrsim 250$,
which eventually undergo fission. As the ejecta becomes less neutron-rich ($Y_e
= 0.25$ and $Y_e = 0.50$), the full r-process is no longer produced; there
are not enough neutrons available per seed nucleus to reach the third peak. At
$Y_e = 0.25$, the first and second r-process peaks are produced. The right panel
of \cref{fig:abundances} shows the final abundances of cases with $Y_e = 0.25$,
$\tau = 7.1\,\text{ms}$, and different initial entropies. Here, the electron
fraction is too high to get to the third r-process peak at most entropies (all
the cases with entropies between 10 and $75\,\kb$ have virtually identical final
abundances as the $s = 10\,\kb$ case). At $s = 100\,\kb$ the third r-process
peak is obtained because the initial composition contains few seed nuclei and alpha
particles are unable to efficiently combine to produce seed nuclei.
Thus, the neutron-to-seed ratio is significantly enhanced.

Animations of the full nucleosynthesis calculations for all seven cases shown in
\cref{fig:abundances} are available at \web. \Cref{fig:movie_frame} shows a
frame from one of the animations.

\subsection{Lanthanide turnoff and heating rate as a
{function~of~\texorpdfstring{$Y_e$}{Ye}}}
\label{sec:hires_ye}

\newcommand{\smAfin}{\smash{$\bar{A}_\text{fin}$}}

\begin{figure*}
\includegraphics[width=\textwidth]{%
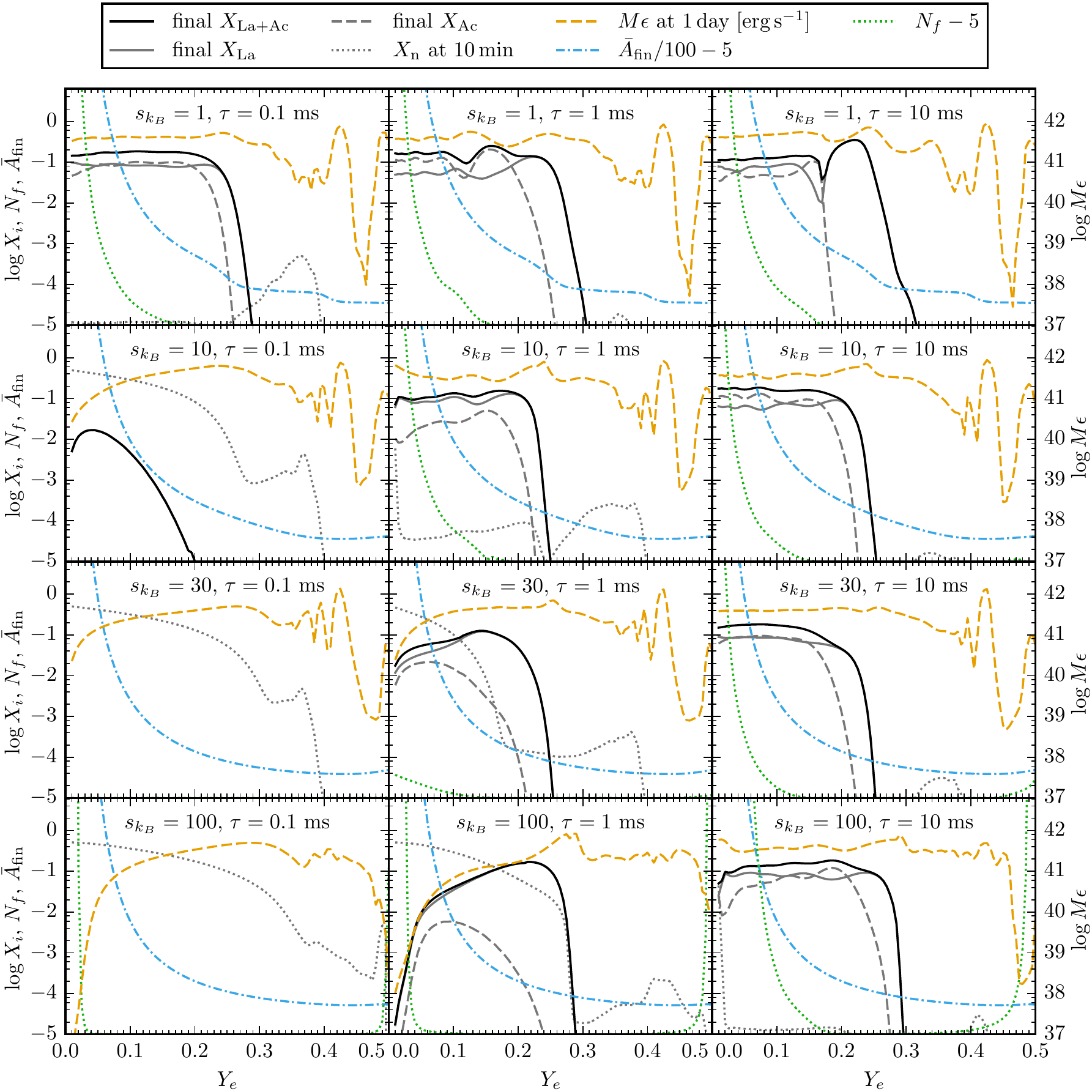}
\caption[]{Results of the high-resolution $Y_e$ runs. The lanthanide and
actinide mass fractions, $X_\text{La}$ and $X_\text{Ac}$, and their sum,
$X_\text{La+Ac}$, are fairly constant up to some critical value of $Y_e$ in most
cases because of fission cycling. The neutron abundance $X_\text{n}$ at
$10\,\text{minutes}$ (the mean lifetime of a free neutron) is an
indicator for a neutron-rich freeze-out, which occurs at high initial entropies
and short expansion timescales, where the neutrons do not have time to capture
on the seed nuclei. The heating rate $M\epsilon$ at $1\,\text{day}$ with $M =
10^{-2}\,M_\odot$ is fairly insensitive to $Y_e$, except at high electron
fractions ($Y_e \gtrsim 0.4$) where some individual nuclides start to dominate
the heating. The
estimated final average mass number \smAfin\ falls off monotonically
with $Y_e$ in all cases except $s = 100\,\kb$, where it rebounds at $Y_e$ very
close to 0.5.  There, the number of seed nuclei decreases drastically because
$\alpha$-particles are initially produced in higher quantities,
which increases the neutron-to-seed ratio. In those cases, the predicted number
of fission cycles $N_f$ is artificially increased at high $Y_e$, because of
production of seed nuclei by the triple-$\alpha$ process. Where equation
\ref{eq:fission_cycle} accurately predicts the number of fission cycles, $N_f$
falls off rapidly with $Y_e$ and the point where it becomes zero is correlated
with the actinide turnoff, because actinides are at the low end of the
fissionable material mass range. Note that we plot \smAfin\ and $N_f$ on linear
scales rather than log scales as all the other quantities. Also, we added a
negative offset of 5 to both \smAfin\ and $N_f$ and we scaled \smAfin\ by
\smash{$1/100$} so that they fit onto our left vertical axis.}
\label{fig:hires_ye}
\end{figure*}

\begin{table}
\caption{$\bar{A}_\text{fin}$ and $Y_e$ at Lanthanide and Actinide Turnoff}
\label{tab:Afin}
\centering
\begin{tabularx}{\linewidth}{c@{\ \,}c@{}X@{}%
>{\centering\arraybackslash}p{0.14\linewidth}>{\centering\arraybackslash}%
p{0.14\linewidth}@{}X@{}%
>{\centering\arraybackslash}p{0.11\linewidth}>{\centering\arraybackslash}%
p{0.11\linewidth}}
\toprule
& &&
\multicolumn{2}{@{\,}c@{\,}}{Lanthanide turnoff\tablenotemark{a}} &&
\multicolumn{2}{@{\,}c@{\,}}{Actinide turnoff\tablenotemark{a}} \\
\cmidrule{4-5} \cmidrule{7-8}
$s$ & $\tau$ && $Y_e$ & \smAfin && $Y_e$ & \smAfin \\[-9pt]
($\kb$) & (ms) & & & & \\
\midrule
   \phn\phn1.0 &       \phn0.1 && 0.27 & \mbox{\phn94} && 0.25 &  \mbox{123}\\
   \phn\phn1.0 & \phn1\phd\phn && 0.28 & \mbox{\phn91} && 0.24 &  \mbox{137}\\
   \phn\phn1.0 &    10\phd\phn && 0.28 & \mbox{\phn93} && 0.18 &  \mbox{192}\\
   \phn\phn1.8 &       \phn0.1 && 0.25 &    \mbox{106} && 0.21 &  \mbox{123}\\
   \phn\phn1.8 & \phn1\phd\phn && 0.27 &    \mbox{100} && 0.21 &  \mbox{125}\\
   \phn\phn1.8 &    10\phd\phn && 0.27 & \mbox{\phn98} && 0.17 &  \mbox{170}\\
   \phn\phn3.0 &       \phn0.1 && 0.23 &    \mbox{118} && 0.20 &  \mbox{135}\\
   \phn\phn3.0 & \phn1\phd\phn && 0.25 &    \mbox{111} && 0.21 &  \mbox{130}\\
   \phn\phn3.0 &    10\phd\phn && 0.27 &    \mbox{106} && 0.18 &  \mbox{150}\\
   \phn\phn5.6 &       \phn0.1 && 0.22 &    \mbox{135} && 0.14 &  \mbox{196}\\
   \phn\phn5.6 & \phn1\phd\phn && 0.23 &    \mbox{127} && 0.21 &  \mbox{138}\\
   \phn\phn5.6 &    10\phd\phn && 0.24 &    \mbox{124} && 0.21 &  \mbox{140}\\
\phn10\phd\phn &       \phn0.1 && 0.13 &    \mbox{223} &&  $-$ &         $-$\\
\phn10\phd\phn & \phn1\phd\phn && 0.24 &    \mbox{121} && 0.21 &  \mbox{139}\\
\phn10\phd\phn &    10\phd\phn && 0.24 &    \mbox{120} && 0.21 &  \mbox{139}\\
\phn18\phd\phn &       \phn0.1 &&  $-$ &           $-$ &&  $-$ &         $-$\\
\phn18\phd\phn & \phn1\phd\phn && 0.24 &    \mbox{102} && 0.20 &  \mbox{130}\\
\phn18\phd\phn &    10\phd\phn && 0.24 &    \mbox{102} && 0.21 &  \mbox{125}\\
\phn30\phd\phn &       \phn0.1 &&  $-$ &           $-$ &&  $-$ &         $-$\\
\phn30\phd\phn & \phn1\phd\phn && 0.24 & \mbox{\phn93} && 0.18 &  \mbox{132}\\
\phn30\phd\phn &    10\phd\phn && 0.24 & \mbox{\phn93} && 0.20 &  \mbox{113}\\
\phn56\phd\phn &       \phn0.1 &&  $-$ &           $-$ &&  $-$ &         $-$\\
\phn56\phd\phn & \phn1\phd\phn && 0.24 & \mbox{\phn94} && 0.16 &  \mbox{143}\\
\phn56\phd\phn &    10\phd\phn && 0.24 & \mbox{\phn94} && 0.21 &  \mbox{109}\\
   100\phd\phn &       \phn0.1 &&  $-$ &           $-$ &&  $-$ &         $-$\\
   100\phd\phn & \phn1\phd\phn && 0.28 & \mbox{\phn94} && 0.18 &  \mbox{148}\\
   100\phd\phn &    10\phd\phn && 0.29 & \mbox{\phn92} && 0.26 &  \mbox{102}\\

\bottomrule
\end{tabularx}
\begin{tabularx}{\linewidth}{>{\raggedright\arraybackslash}X}
${}^\text{a}$ Turnoff is when the mass fraction $X_\text{La}$ or $X_\text{Ac}$
drops below $10^{-3}$. The columns show the maximum $Y_e$ and corresponding
minimum \smAfin\ for which $X_i \geq 10^{-3}$. A dash ($-$) denotes
that $X_i < 10^{-3}$ for all $Y_e$, which means there is a neutron-rich
freeze-out.
\end{tabularx}
\end{table}

\Cref{fig:hires_ye} shows the final lanthanide and actinide mass fractions
$X_\text{La}$ and $X_\text{Ac}$, respectively,
along with the neutron mass fraction $X_\text{n}$ at $10\,\text{minutes}$,
which is the mean lifetime of a free neutron. Also
shown is \smAfin, which is an estimate of the final average mass
number $A$ of the material. It is defined as
\begin{align}
\label{eq:Abarfin}
\bar{A}_\text{fin} = \frac{1}{Y_\text{seed}(0) + Y_\alpha(0)/18}\,,
\end{align}
where $Y_\alpha(0)$ is the initial $\alpha$-particle abundance and
$Y_\text{seed}(0)$ is the initial seed abundance (sum of abundances of all
nuclides with $A \geq 12$). Since the
$\alpha$-process ceases around Kr in neutron rich conditions \citep{woosley:92},
it takes around eighteen $\alpha$ particles to make a seed nucleus.
Therefore,
the quantity in the denominator of \cref{eq:Abarfin} is approximately
the number abundance of heavy nuclei present at the end of the r-process.  We
then
arrive at \cref{eq:Abarfin} by assuming that the total mass fraction of
heavy nuclei at the end of the calculation is unity. Clearly,
this assumption
breaks down if there is fission cycling, because then the number of seeds at
the end is much larger than the number of initial seeds plus those produced
by the $\alpha$-process. However, we are interested in the value of
\smAfin\ at the actinide and lanthanide turnoff, which preclude
significant fission cycling because fission cycling only happens if nuclides
heavier than actinides are produced, and so there is no problem in using the
definition
in \cref{eq:Abarfin}. At low electron fractions, $\alpha$-rich freeze-out
does not occur due to the low initial abundance of $\alpha$ particles. We
emphasize that \smAfin\ only depends on
the initial abundances, and thus it is useful to determine whether a
certain trajectory is likely to produce large quantities of lanthanides or
actinides, without having to perform any nucleosynthesis calculation.

\Cref{tab:Afin} shows the values of $Y_e$ and \smAfin\ at which
lanthanide and actinide production ceases (mass fraction goes below
\smash{$10^{-3}$}). In other words, if $Y_e$ is lower than or
\smash{\smAfin}
larger than what is shown in \cref{tab:Afin}, then \smash{$X_\text{La} \geq
10^{-3}$} or
\smash{$X_\text{Ac} \geq 10^{-3}$}. The lanthanide turnoff is at
\smash{$\bar{A}_\text{fin}
\sim 100$} and the actinide turnoff is at \smash{$\bar{A}_\text{fin}\sim 130$}.
The cases
where \smash{$X_\text{La} < 10^{-3}$} or \smash{$X_\text{Ac} < 10^{-3}$} for
all $Y_e$ are
denoted by ``$-$'' in \cref{tab:Afin}, and they correspond to the strong
neutron-rich freeze-outs in \cref{fig:hires_ye}, which means that the r-process
did not happen (or at least not efficiently) in those cases because after about
$10\,\text{min}$ we are just left with free neutrons that will now decay to
protons. In the case $s10\tau0.1$ (which
stands for $s = 10\,\kb$ and $\tau = 0.1\,\text{ms}$) where
lanthanides are made, but no actinides above a mass fraction of \smash{$10^{-3}$}, we
see a
weaker neutron-rich freeze-out in \cref{fig:hires_ye}. The neutron-rich
freeze-outs happen at high initial entropies and short expansion timescales,
where the ejecta is very hot and expands quickly, which leaves little time for
neutrons to capture on seed nuclides.  There is also a neutron-rich freeze-out in
$s30\tau1$ and $s100\tau1$ models, but the freeze-out is weak enough to allow
lanthanides and actinides to be produced, albeit in lower quantities.
\citet{metzger:14b} suggested that a kilonova containing some mass with such
short dynamical timescales could be preceded by an ultraviolet transient
powered by these frozen-out neutrons.

\Cref{fig:hires_ye} shows that the heating rate from decay at $1\,\text{day}$ is
quite insensitive to $Y_e$ at $Y_e \lesssim 0.35$ and also fairly insensitive
to the amount of lanthanides and actinides produced. As long as
$X_\text{La+Ac}$ is more or less constant as a function of $Y_e$, $M\epsilon$ at
$1\,\text{day}$ is also fairly constant.  When the lanthanides turn off,
there is a small bump in the heating rate in most cases and at larger $Y_e$,
after lanthanides have completely gone away, the
heating rate drops only slightly (an order of magnitude or less). One might
expect a larger decline of the heating rate once the full r-process stops
happening, because the material is less neutron-rich overall, more stable nuclei
are produced directly, and thus the total
radioactive decay energy should be lower. This is indeed true and we
verified it by looking at the
integrated nuclear heating amount as a function of $Y_e$ (for fixed $s$ and
$\tau$). We find that in most cases the total amount of heating drops by 1.5
to 2
orders of magnitude as $Y_e$ goes from low values to high values. There is a
smaller drop in the heating rates shown in \cref{fig:hires_ye}, because
there we only plot the instantaneous heating rate at $1\,\text{day}$. Since
the $\beta$-decay energy is correlated with the decay timescale, we always see a
similar instantaneous decay rate at the same point in time, as long as we have a
collection of nuclides with half-lives at around a day. The picture changes at
$Y_e \gtrsim 0.35$ because there the final composition is dominated by one or a
few individual
nuclides, as opposed to a large ensemble of nuclides, which then determine the
heating rate. This is discussed in
detail in \sref{sec:lan_prod}.

Since our parameter space is three-dimensional, we can go beyond giving a
simple $Y_e$ cutoff for lanthanide production. We
use a heuristic method to fit for the coefficients of three inequalities
in $Y_e$, $\ln s$, and $\ln \tau$ that separate the lanthanide-rich and
lanthanide-free regions of the parameter space. We find that
\vspace{-1ex}
\begin{align*}
\def\arraystretch{1.2}
\begin{array}{@{}r@{\,}c@{\,}r@{\,}c@{\,}r@{\,}c@{\,}r@{}}
\multicolumn{7}{@{}c@{}}{\text{$X_\text{La+Ac} \geq 10^{-3}$ if and only if}}
\\[2pt]
-1.00\,Y_e & - & 0.00744 \ln s_{k_B} & + & 0.000638 \ln
\tau_\text{ms}& + & 0.259
\geq 0\phantom{,} \\
\multicolumn{7}{@{}c@{}}{\text{and}} \\
-0.990\,Y_e & + & 0.117\ln s_{k_B} & - & 0.0783\ln\tau_\text{ms} & + & 0.452
\geq 0\phantom{,} \\
\multicolumn{7}{@{}c@{}}{\text{and}} \\
-0.799\,Y_e &  - & 0.288 \ln s_{k_B} & + & 0.528 \ln \tau_\text{ms}& + & 1.88
\geq 0,
\end{array}
\end{align*}
\vspace{-1ex} \par
\noindent where $s_{k_B}$ is the entropy $s$ in units of $\kb$ and
$\tau_\text{ms}$ is the expansion timescale $\tau$ in units of milliseconds. The
above statement only fails for 97 out of 4913 points in our parameter space,
i.e.\ it is true for 98\% of the parameter space. Most of the points where the
above fails are very close to one of the planes, but there are a few points
further away from the boundaries that fail too. Those points are all at very low
$Y_e$, high entropy, and very short expansion timescale, where we get strong
neutron-rich freeze-out. The results of the full parameter space are discussed
in detail in \sref{sec:lan_prod}.

\subsection{Fission cycling}

If the r-process is strong enough to produce nuclides with masses near 300,
these nuclides fission and the fission products then capture more neutrons,
eventually getting up to $A\sim 300$ and fissioning again, creating a fission
cycle. Thus fission cycling limits the maximum mass of nuclides produced in the
r-process, which washes out the initial conditions of the ejecta and hence the
final abundances are determined by nuclear physics rather than the properties of
the outflow.

The quantity $N_f$ shown in \cref{fig:hires_ye} is an estimate for the number of
fission cycles that occurred during nucleosynthesis. It is defined as
\begin{align}
\label{eq:fission_cycle}
N_f = \frac{Y_\text{seed}(t = t_n)}{Y_\text{seed}(t = 0)} - 1,
\end{align}
where $Y_\text{seed}(t = t_n)$ is the abundance of all seed nuclides ($A \geq
12$)
at the time that neutrons are exhausted (when $X_n \leq 10^{-4}$) and
$Y_\text{seed}(t = 0)$ is the initial abundance of seed nuclei. This estimate
for the number of fission cycles rests on the assumption that only fission can
create additional seed nuclides. When a neutron captures on a seed nuclide, it
creates a heavier nuclide, but it will not increase the total number (and
hence abundance) of seed nuclides in the ejecta. However, if a heavy nuclide
(which is counted as a seed nuclide) fissions, then there are two seed
nuclides in its place. Thus
comparing the number of heavy nuclides at the time when neutron
capture ceases to the initial number of heavy nuclides tells us how
many additional heavy nuclides were produced. For example, if $Y_\text{seed}(t
= t_n) = Y_\text{seed}(t =0)$, then no additional heavy nuclides were produced
and thus there was no fission cycling, hence $N_f = 0$. But if $Y_\text{seed}(t
= t_n) = 3Y_\text{seed}(t =0)$, for example, then (on average) each initial
heavy nuclide produced two additional heavy nuclides and so there were two
fission cycles, hence $N_f = 2$. Note that this method of estimating the
number of fission cycles breaks down
if nuclides with $A \geq 12$ are produced from nuclides with $A < 12$, e.g.\
\smce{^{12}C} from three \smce{^{4}He}. This happens most prominently at $Y_e$
close to 0.5 and at high entropies, where fission will clearly not occur.

\begin{figure*}
\includegraphics[width=\textwidth]{%
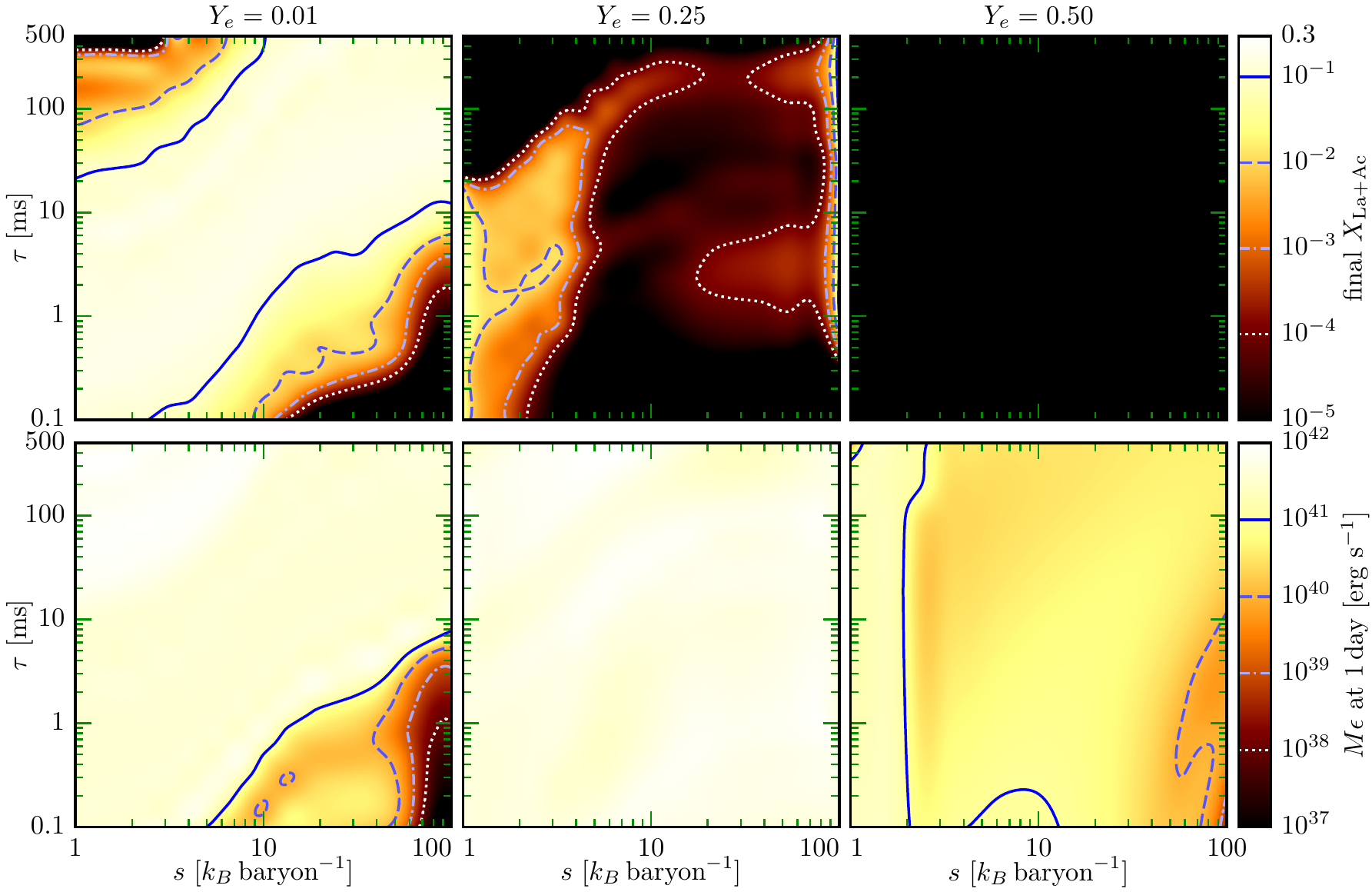}
\caption[]{Slices of constant electron fraction showing the lanthanide and
actinide mass fraction $X_\text{La+Ac}$ and the heating rate $M\epsilon$ at
$1\,\text{day}$ with \smash{$M = 10^{-2}\,M_\odot$}. For $Y_e = 0.01$, the
high-$s$/small-$\tau$ corner is lanthanide-free because the high entropy
produces very light seed nuclides, fewer seed nuclei are produced due to an
$\alpha$-rich freeze-out, and neutron capture begins at low density due
to the high entropy (see the text for more discussion).  The
low-$s$/large-$\tau$ corner is lanthanide-free because the slow expansion
timescale results in significant late-time heating, which drives the ejecta back
to \NSE{}, but at those late times, $\beta$-decays have significantly raised the
electron fraction and so the r-process starts again but at a much higher $Y_e$,
which does not produce lanthanides. The $Y_e = 0.25$ slice is the transition
between lanthanide-rich and lanthanide-free. At low entropies we can still make
significant amounts of lanthanides because the seed nuclides are heavy, and at
very high entropies we initially have a lot of free neutrons and $\alpha$
particles, which can produce significant amounts of heavy elements. Finally, at
$Y_e = 0.50$ the material is simply not neutron-rich enough to make any
lanthanides. The heating rate at $1\,\text{day}$ is quite insensitive to $s$ and
$\tau$, except at low $Y_e$, where it is significantly smaller at high entropies
and fast expansion timescales because a neutron-rich freeze-out happens. The
uniformity in the heating rate is due to the fact that there is an
ensemble of nuclides contributing to the heating. And since we are considering
the heating at $1\,\text{day}$, we tend to pick up nuclides with
similar decay energies (because the decay energy is correlated with the
half-life), leading to similar heating rates even if the composition varies.}
\label{fig:ye_slices}
\end{figure*}

As expected, there are many fission cycles at low $Y_e$ where large amounts of
lanthanides and actinides are produced. In the regions with significant fission
cycling, $X_\text{La}$, $X_\text{Ac}$, and $\epsilon$ are fairly insensitive to
$Y_e$ because fission cycling effectively limits the maximum mass of nuclides
that are produced to $A\sim 300$. As the ejecta becomes less neutron-rich,
fewer fission cycles occur because there are not enough free neutrons to produce
fissionable material with $A \gtrsim 250$.

In most panels in \cref{fig:hires_ye} we see that the production of actinides is
closely tied to fission cycling; actinides go away just after fission cycling
stops. If the r-process cannot get to $A\sim 250$, it cannot create actinides
and it cannot create fissionable material. Furthermore, in most panels, but
especially in $s1\tau1$ and $s1\tau10$ there is an increase in $X_\text{Ac}$ and
decrease in $X_\text{La}$ at the electron fraction where fission cycling stops and
just before actinides are not produced. Just as fission cycling stops, the
r-process can get to about $A=250$, but not much above. This means that
actinides can still be produced, but they are not being fissioned (because only
lighter actinides are produced or there are no more free neutrons to initiate
fission). Lanthanides have a mass around 150 and so they can be created from
fission products.  When fission is just turning off, we lose a small source
of lanthanides leading to the (small) decline in $X_\text{La}$ that can be
prominently seen in $s1\tau10$ in \cref{fig:hires_ye} at $Y_e = 0.17$.

\begin{figure*}
\includegraphics[width=\textwidth]{%
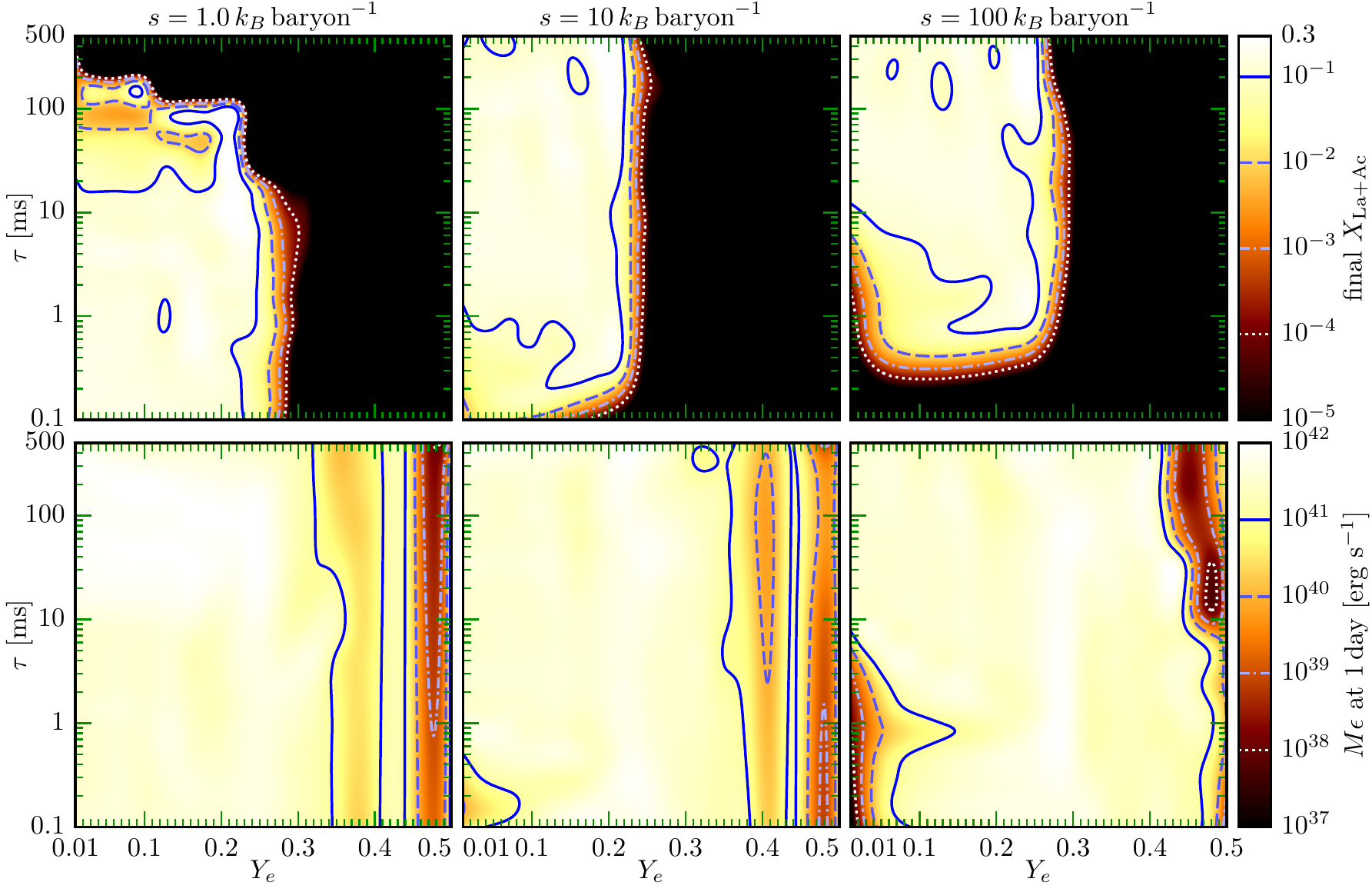}
\caption[]{Slices of constant entropy showing the lanthanide and actinide mass
fraction $X_\text{La+Ac}$ and the heating rate $M\epsilon$ at $1\,\text{day}$
with \smash{$M = 10^{-2}\,M_\odot$}. At $s=1\,\kb$, no lanthanides are produced
at large
expansion timescales because the material heats up significantly at late times,
which restarts the r-process at late times after $Y_e$ has risen to about 0.3.
At $s=100\,\kb$, no lanthanides are produced when the dynamical timescale is
short for the reasons discussed in the caption of figure \ref{fig:ye_slices}.
In all cases, there is a critical value of $Y_e$
where lanthanide production abruptly ceases. The heating rate at $1\,\text{day}$
only shows some structure at high $Y_e$ where certain individual nuclides
dominate the heating. The reduced heating in the low-$Y_e$/small-$\tau$ corner
of $s = 100\,\kb$ is due to a neutron-rich freeze-out that occurs there.}
\label{fig:s_slices}
\end{figure*}

\begin{figure*}
\includegraphics[width=\textwidth]{%
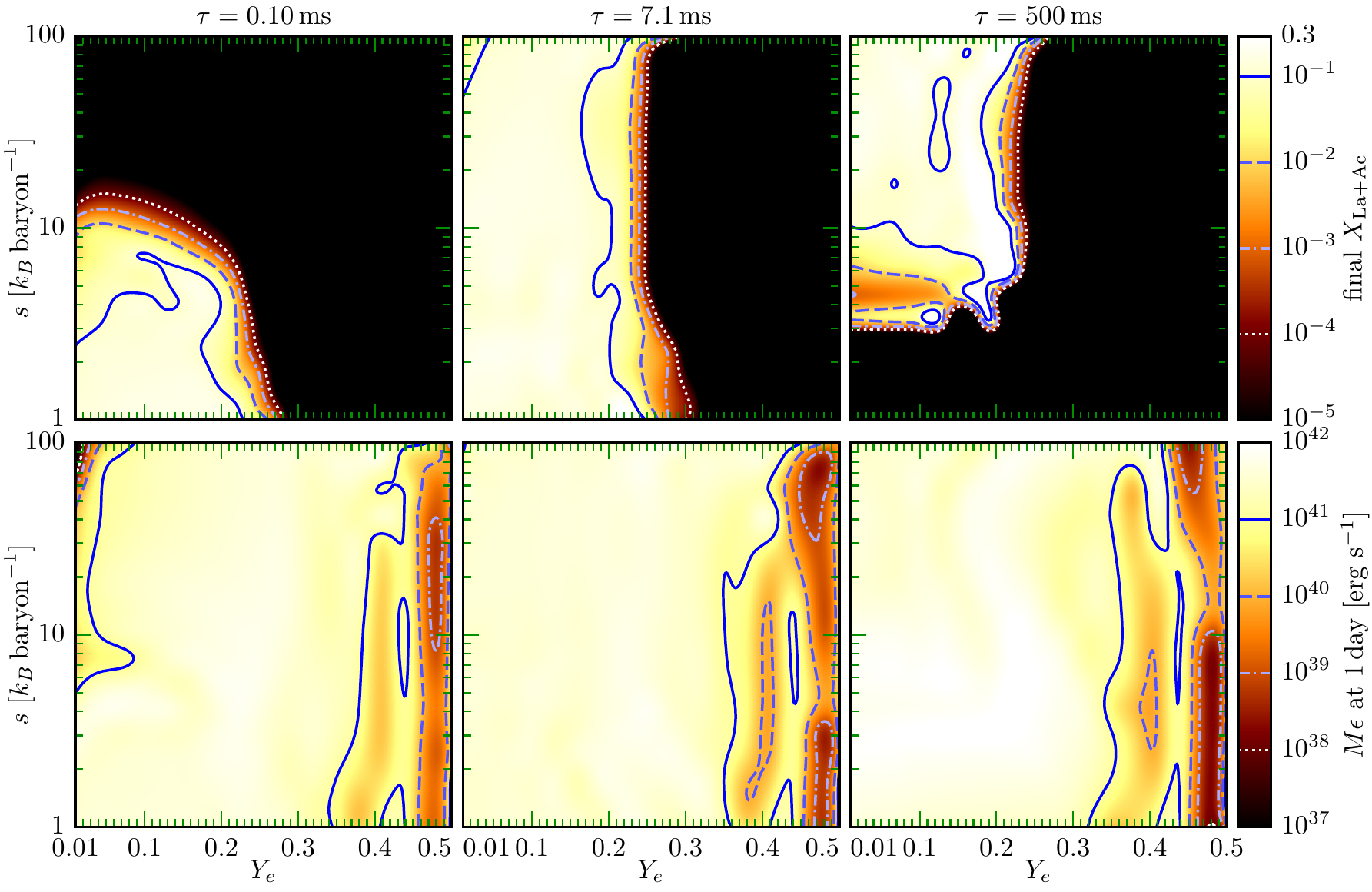}
\caption[]{Slices of constant expansion timescale showing the lanthanide and
actinide mass fraction $X_\text{La+Ac}$ and the heating rate $M\epsilon$ at
$1\,\text{day}$ with \smash{$M = 10^{-2}\,M_\odot$}. At $\tau =
0.10\,\text{ms}$, there
are no lanthanides at high entropies because the neutrons have no time to
capture on the light seed nuclides. At $\tau=500\,\text{ms}$, there are no
lanthanides at low entropies because the heavy, neutron-rich seed nuclei lead to
substantial late-time heating, which restarts the r-process at $Y_e \sim 0.3$,
which is not neutron-rich enough to produce lanthanides. In all cases, there is
a fairly uniform lanthanide cutoff as $Y_e$ goes beyond a critical value. The
heating rate at $1\,\text{day}$ only shows structure at high $Y_e$ where certain
individual nuclides dominate the heating.}
\label{fig:tau_slices}
\end{figure*}

\subsection{Lanthanide production and heating rate in the full parameter space}
\label{sec:lan_prod}

Since the amount of lanthanides determines the opacity of the ejecta and the
nuclear heating rate determines the amount of energy available for the
electromagnetic transient, we are especially interested in how these two
quantities are correlated in our parameter space.
\Cref{fig:ye_slices,fig:s_slices,fig:tau_slices} show slices of the final
lanthanide
and actinide mass fractions, $X_\text{La+Ac}$,
and heating rates at $1\,\text{day}$ for the extreme and intermediate values of
$Y_e$, $s$, and $\tau$. All the other slice plots are available at \web. In the
following, the term ``lanthanide'' will stand for both lanthanides and
actinides, unless actinides are specifically mentioned.
Unsurprisingly, $X_\text{La+Ac}$ depends most strongly on $Y_e$ and the ejecta
is lanthanide-free for $Y_e \gtrsim 0.26$. However, even for a very low $Y_e$ of
0.01, there are some combinations of $s$ and $\tau$ that yield a lanthanide-free
ejecta (see upper left panel of \cref{fig:ye_slices}). Specifically, at high
entropies ($s \gtrsim 20\,\kb$) and small expansion timescales ($\tau \lesssim
1\,\text{ms}$), no lanthanides are produced. The reason for this is that
neutron capture begins at a
lower density because of the high entropy (for a fixed temperature at which
neutron capture begins) and therefore the neutron capture timescale is
increased.  This---in combination with light seed nuclei, a large initial
neutron abundance, a potentially $\alpha$-rich freeze-out, and a short dynamical
timescale---prevents production of lanthanides and sometimes results in a
neutron-rich freeze-out.  At lower entropies, the seed
nuclei are heavier and the density is higher during the neutron capture period,
allowing neutrons to capture on them even at small expansion
timescales. And at larger expansion timescales, there is more time for the
neutrons to capture on the light seed nuclei even at very high entropies. This
is reflected in the upper right panel of \cref{fig:s_slices} where no
lanthanides are produced at small expansion timescales at $s = 100\,\kb$, and in
the upper left panel of \cref{fig:tau_slices} where no lanthanides are produced
at high entropies at $\tau = 0.1\,\text{ms}$.

There is another lanthanide-free corner in the upper left panel of
\cref{fig:ye_slices} at very large expansion timescales ($\tau\gtrsim
400\,\text{ms}$) and low entropies ($s\lesssim 3\,\kb$). Here, the full
r-process
is being made, since the material is very neutron-rich, but because the
expansion timescale is so long, the density is still quite high (about
\smash{$10^{10}\,\text{g}\,\text{cm}^{-3}$}) when neutron burning ceases. All
the heavy
elements then decay and considerably heat up the material (to above
$7\,\text{GK}$), which destroys all heavy nuclides via photodissociation and
brings the composition back to \NSE{}. Only after tens of seconds has the
material cooled down enough for neutron captures to happen again, but by then,
$\beta$-decays have raised $Y_e$ to about 0.3. Thus we now get an r-process with
an initial $Y_e$ of 0.3, which is not neutron-rich enough to produce
lanthanides. At faster expansion rates (smaller $\tau$) the density falls off
faster, resulting in less dramatic heating that cannot force the composition
into \NSE{}. Because we obtain the initial density from solving for \NSE{} at
the prescribed entropy, $Y_e$, and $T = 6\,\text{GK}$, the initial density is
lower at higher  entropies ($s \gtrsim 3\,\kb$) and so even though the density
remains close to the initial value at $\tau = 500\,\text{ms}$, the density is
not high enough to produce heating that results in \NSE{}. This is reflected in
the upper left panel of \cref{fig:s_slices} where the ejecta is lanthanide-free
at large expansion timescales at $s = 1\,\kb$, and in the upper right panel of
\cref{fig:tau_slices} where no lanthanides are produced at low entropies at
$\tau = 500\,\text{ms}$.

The $Y_e = 0.25$ slice in \cref{fig:ye_slices} is right at the transition from
lanthanide-rich to lanthanide-free ejecta. The upper panels of
\cref{fig:s_slices,fig:tau_slices} show clearly that this transition is very
sharp at $Y_e \sim 0.22 - 0.30$. In the upper middle panel of
\cref{fig:ye_slices}, the low-$s$/large-$\tau$ corner that is lanthanide-free
has expanded and so has the high-$s$/small-$\tau$ corner, relative to the
$Y_e=0.01$ panel.  Additionally, lanthanide production is suppressed at
intermediate entropies ($5\,\kb \lesssim s \lesssim 90\,\kb$). At low entropies,
we still get an r-process because the seed nuclei are very heavy and thus
require fewer neutrons to capture on them to make the r-process distribution.
At very high entropies, the initial composition includes a large fraction of
free neutrons and $\alpha$ particles.  At high entropies, production of seed
nuclei via neutron catalyzed triple-$\alpha$ is suppressed \citep{hoffman:97},
which
reduces the number of seed nuclei and thereby increases the neutron-to-seed
ratio.  These conditions allow for the production of the r-process nuclei. With
$Y_e \gtrsim 0.3$, lanthanides are not produced at any entropy and expansion
timescale combination, since the ejecta is not neutron-rich enough. In
\sref{sec:hires_ye} we discussed in detail how the final
lanthanide and actinide mass fractions depend on $Y_e$.

The lower row of panels in \cref{fig:ye_slices,fig:s_slices,fig:tau_slices}
shows the heating rate (actually $M\epsilon$ where $M = 10^{-2}\,M_\odot$) at
$1\,\text{day}$.  For $0.04 \lesssim Y_e \lesssim 0.35$ all the $Y_e$ slices are
very similar to the lower middle panel of \cref{fig:ye_slices}, with virtually
no
structure. At $Y_e = 0.01$, the high-$s$/small-$\tau$ corner has significantly
less heating because the initial density is very low ($\rho_0 \sim 8\times
10^5\,\text{g}\,\text{cm}^{-3}$) and this, coupled with the rapid expansion
timescale ($\tau = 0.1\,\text{ms}$) and the fact that the initial composition
contains few seed nuclei (98\% of the mass is neutrons), means there is little
opportunity for neutron capture. For larger expansion timescales, the initial
conditions remain the same (low initial density and 98\% of the mass is
neutrons), but because the density decreases more slowly, there is sufficient
time for neutrons to capture on the few seed nuclei available and make a full
r-process. At lower initial entropies, the initial density is larger (e.g.\
$4\times 10^6\,\text{g}\,\text{cm}^{-3}$ at $s = 32\,\kb$) so that the density
remains higher even with a rapid expansion, giving the neutrons a better chance
to capture on seed nuclei---of which there are slightly more available---leading
to a moderate r-process. This is reflected in the low-$Y_e$/small-$\tau$ corner
of the lower right panel in \cref{fig:s_slices} and in the low-$Y_e$/high-$s$
corner of the lower left panel in \cref{fig:tau_slices}.

For $Y_e \gtrsim 0.35$ we start to see large variations in the heating rate at
$1\,\text{day}$ as a function of $Y_e$, which can be seen in all lower panels
in
\cref{fig:s_slices,fig:tau_slices}. But the heating is still quite insensitive
to $s$ and $\tau$, as the lower right panel of \cref{fig:ye_slices} shows. This
variation as a function of $Y_e$ at high $Y_e$ can also be seen in
\cref{fig:hires_ye}. There is a pronounced peak in the heating rate at
$1\,\text{day}$ at $Y_e = 0.425$ in all but the $s = 100\,\kb$ cases. This peak
is due to the decay of \smce{^{66}Cu} (half-life of $5\,\text{minutes}$) which
comes from the decay of \smce{^{66}Ni}, which has a half-life of
$55\,\text{hours}$. \smce{^{66}Ni} has 28 protons and 38 neutrons and so its
electron fraction is $28/66 \approx 0.424$, which is very close to $Y_e =
0.425$, the initial electron fraction of the material. Thus the initial
\NSE{} distribution contains a larger quantity of \smce{^{66}Ni} at $Y_e =
0.425$
than at different $Y_e$, which leads to excessive heating via the decay
chain described above because \smce{^{66}Cu} has a fairly large $Q$-value of
$2.6\,\text{MeV}$. At $s = 100\,\kb$ the initial neutron-to-seed ratio is much
larger than at lower entropies and so significant neutron burning occurs even at
high $Y_e$, which washes out the strong dependence of the heating rate at
$1\,\text{day}$ on $Y_e$.

In \cref{fig:hires_ye}, there are also large minima in the heating rate at
$1\,\text{day}$ in all but the $s = 100\,\kb$ cases at electron fractions
between 0.45 and 0.48, depending on $s$ and $\tau$. These minima can also be
seen in \cref{fig:s_slices,fig:tau_slices}. In those cases, \NSE{}
preferentially produces stable isotopes in the initial composition, which
drastically reduces the heating. For example, the cases with $s = 1\,\kb$ have
the minima at $Y_e = 0.465$ and over 80\% of the initial mass is either stable
or has a half-life of more than $100\,\text{days}$. The most abundant nuclide
(37\% of the mass) is \smce{^{56}Fe}, which is stable and has $Y_e = 26/56
\approx 0.464$, which is why the minimum occurs at $Y_e = 0.465$, because that
favors \smce{^{56}Fe} the most. As another example, the $s = 10\,\kb$ cases have
the minima at $Y_e = 0.45$, where \smce{^{58}Fe} and \smce{^{62}Ni} are
preferentially produced by \NSE{}, which have electron fractions of 0.448 and
0.452, respectively.

As in \sref{sec:hires_ye}, we do not find a significant correlation between the
amount of lanthanides and actinides produced with the heating rate at
$1\,\text{day}$. The heating rate at $1\,\text{day}$ is very uniform at values
of $Y_e$ where lanthanides are produced. Since we are looking at the heating
rate at a specific time, we will always pick out the nuclides with a half-life
of about $1\,\text{day}$ (or decay products of nuclides that decay on a one-day
timescale). Because the decay energy is correlated with the half-life and
because we always have a collection of different nuclides, we obtain roughly the
same heating rate at $1\,\text{day}$ regardless of the exact composition of the
ejecta. This is no longer true at higher $Y_e$, where the composition can be
dominated by individual nuclides, which then determine the heating rate.

\subsection{Fitted nuclear heating rates}

\begin{figure}
\includegraphics[width=\linewidth]{%
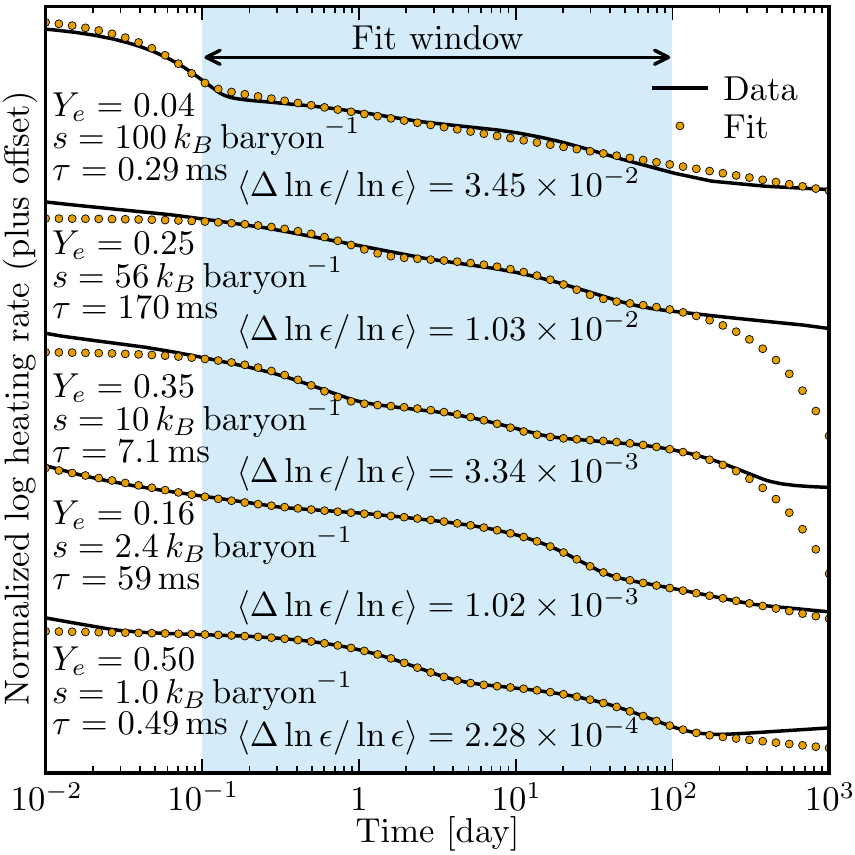}
\caption[]{Some heating rate fits showing the fits with the largest and smallest
error, and fits with errors in between. The heating rate is only fitted inside
the
fit window (0.1 to $100\,\text{days}$). We use a power law with up to two
exponential terms, or up to three exponential terms without a power law show in
\cref{eq:heating_fit}, whichever produces the best fit. The fit error
$\langle\Delta\ln\epsilon/\ln\epsilon\rangle$ is defined in
\cref{eq:mean_frac_log_error}. As the second and third case from the top show,
the fit can be quite bad outside the fit window. This is no surprise since we
do not fit the data outside the fit window and because we only use up to three
exponential terms. In reality, there are hundreds of individual nuclides
contributing to the total heating rate and each one contributes a different
exponential term.}
\label{fig:heating_rate_fits}
\end{figure}

For each nucleosynthesis calculation, we calculate a parametric fit for the
nuclear heating rate $\epsilon(t)$ between 0.1 and $100\,\text{days}$ (the fit
window). The fit has the form
\begin{align}
\hat\epsilon(t) = A t^{-\alpha} + B_1 e^{-t/\beta_1}  + B_2 e^{-t/\beta_2} +
B_3 e^{-t/\beta_3}, \label{eq:heating_fit}
\end{align}
where $t$ and $\beta_i$ are in days and $\hat\epsilon(t)$ is in
$\text{erg}\,\text{s}^{-1}\,\text{g}^{-1}$. We use at most six parameters for
the fit, so either $A$ and $\alpha$ are zero or one or more of $B_i$ and
$\beta_i$ are zero. We use a weighted fit where the range 0.1 to
$100\,\text{days}$ has a weight of one and the weight decreases linearly to zero
in logspace from 0.1 to $0.05\,\text{days}$ and from 100 to $200\,\text{days}$.
We use a heuristic method to find the global best
fit for all six types of fits (power law with 0, 1, or 2 exponentials, or 1, 2,
or 3 exponentials without a power law term). The best of these six fits is then
selected with a small penalty term for the number of parameter pairs. The
fitting error is multiplied by 1.1 for each parameter pair in excess of one, so
that we do not pick up meaningless parameters that improve the fit by less than
10\%.

For consistency, we calculate the fitting error at the same times $t_i$ for all
cases and we interpolate the actual heating rate to those times, which are 500
points uniformly sampled in logspace between $10^{-2}$ and $10^3\,\text{days}$
(however, points before $0.05\,\text{days}$ and after $200\,\text{days}$ have
zero weight and thus do not contribute to the fitting error, as explained
above). The fit error used for finding the optimal fit parameters is the sum of
squares of the log difference, i.e.\
\begin{align}
\text{fit error} = \sum_i w_i \left(\ln\epsilon(t_i) -
\ln\hat\epsilon(t_i)\right)^2,
\end{align}
where $w_i$ is the weight of time $t_i$. This error measure works well for the
optimization algorithm to find the best parameters, but it carries little
physical meaning. To be able to intuitively judge the quality of a particular
fit, we define the mean fractional log error as
\begin{align}
\left\langle\frac{\Delta \ln\epsilon}{\ln\epsilon}\right\rangle =
\left\langle\frac{|\ln\epsilon(t_i) -
\ln\hat\epsilon(t_i)|}{\ln\epsilon(t_i)}\right\rangle,
\label{eq:mean_frac_log_error}
\end{align}
where the average runs over all times $t_i$ such that $0.1\,\text{days} \leq t_i
\leq 100\,\text{days}$. We only fit the total heating rate, but we also
provide the average heating contribution due to fission reactions in the fit
window.

The best and worst heating rate fits, as well as some fits of intermediate
quality, are shown in \cref{fig:heating_rate_fits}. About 80\% of all
high-resolution \verb\sym0\ fits have
$\left\langle\Delta\ln\epsilon/\ln\epsilon\right\rangle \leq 0.5\%$ and about
95\% have a mean fractional log error of at most 1\%. Since we do not include
$\beta$-delayed fission reactions, the heating due to fission in our fit window
(0.1 to $100\,\text{days}$) is solely due to spontaneous fission and it is close
to constant during the fit window because there is usually one nuclide that
dominates the fission heating. In 85\% of all cases it varies by less than a
factor of two within the fit window, and in 99\% of all cases it varies by less
than a factor of three. Thus it is sufficient to report the geometric mean of
the heating rate due to fission over the fit window.  Fits to the heating rates
over our entire parameter space are available at \web.

\subsection{Dominant nuclear decays}

\begin{table*}[t!]
\caption{Average Integrated Fractional Heating Contributions $\bar{f}_i$ of the
High-Resolution \texttt{sym0}\tablenotemark{a} Runs}
\label{tab:dominant_nuclides}
\renewcommand{\arraystretch}{1.2}
\centering
\begin{tabularx}{0.9\textwidth}{%
@{}X@{}lc@{}X@{}%
@{}X@{}lc@{}X@{}%
@{}X@{}lc@{}X@{}%
@{}X@{}lc@{}X@{}%
@{}X@{}lc@{}X@{}}
\toprule
\multicolumn{16}{@{}c}{$Y_e$ Bins\tablenotemark{b}} &
\multicolumn{4}{p{0.16\textwidth}@{}}{\centering Overall\tablenotemark{c}} \\
\cmidrule(r){1-16}
\multicolumn{4}{@{}p{0.16\textwidth}}{\centering$0 < Y_e \leq 0.125$} &
\multicolumn{4}{p{0.16\textwidth}}{\centering$0.125 < Y_e \leq 0.250$} &
\multicolumn{4}{p{0.16\textwidth}}{\centering$0.250 < Y_e \leq 0.375$} &
\multicolumn{4}{p{0.16\textwidth}}{\centering$0.375 < Y_e \leq 0.5$} &
\multicolumn{4}{p{0.16\textwidth}@{}}{\centering($0 < Y_e \leq 0.5$)} \\
\cmidrule(r){1-4} \cmidrule(lr){5-8} \cmidrule(lr){9-12} \cmidrule(lr){13-16}
\cmidrule(l){17-20}
& Nuclide & $\bar{f}_i$ &&& Nuclide & $\bar{f}_i$ &&& Nuclide & $\bar{f}_i$ &&&
Nuclide& $\bar{f}_i$ &&& Nuclide & $\bar{f}_i$ & \\ \midrule
&  \ce{^{132}I} &    22.59\% &&&  \ce{^{132}I} &    26.49\% &&&  \ce{^{89}Sr} &
\phn9.01\% &&&  \ce{^{66}Cu} &    13.21\% &&&  \ce{^{132}I} &    13.99\% & \\
& \ce{^{200}Au} & \phn4.46\% &&&  \ce{^{131}I} & \phn5.52\% &&&  \ce{^{72}Ga} &
\phn5.91\% &&&  \ce{^{57}Ni} &    10.83\% &&&  \ce{^{66}Cu} & \phn4.42\% & \\
& \ce{^{128}Sb} & \phn4.26\% &&& \ce{^{128}Sb} & \phn4.66\% &&&  \ce{^{132}I} &
\phn5.00\% &&&  \ce{^{59}Fe} & \phn7.47\% &&&  \ce{^{89}Sr} & \phn3.51\% & \\
& \ce{^{249}Bk} & \phn4.23\% &&& \ce{^{132}Te} & \phn3.78\% &&&  \ce{^{59}Fe} &
\phn4.77\% &&&  \ce{^{89}Sr} & \phn5.21\% &&&  \ce{^{57}Ni} & \phn3.18\% & \\
& \ce{^{132}Te} & \phn3.22\% &&& \ce{^{125}Sn} & \phn3.37\% &&&  \ce{^{78}As} &
\phn4.65\% &&&  \ce{^{77}As} & \phn4.79\% &&&  \ce{^{59}Fe} & \phn3.04\% & \\
&  \ce{^{131}I} & \phn3.13\% &&&  \ce{^{133}I} & \phn3.06\% &&& \ce{^{125}Sn} &
\phn3.64\% &&&  \ce{^{77}Ge} & \phn4.18\% &&& \ce{^{128}Sb} & \phn2.67\% & \\
& \ce{^{252}Cf} & \phn3.09\% &&& \ce{^{129}Sb} & \phn2.85\% &&& \ce{^{103}Ru} &
\phn3.24\% &&&  \ce{^{61}Cu} & \phn3.20\% &&&  \ce{^{131}I} & \phn2.59\% & \\
&  \ce{^{133}I} & \phn3.09\% &&& \ce{^{127}Sb} & \phn2.79\% &&&   \ce{^{91}Y} &
\phn3.08\% &&&  \ce{^{62}Cu} & \phn3.04\% &&&  \ce{^{78}As} & \phn2.27\% & \\
& \ce{^{202}Au} & \phn2.89\% &&& \ce{^{140}La} & \phn2.56\% &&&  \ce{^{66}Cu} &
\phn2.97\% &&&  \ce{^{56}Ni} & \phn3.00\% &&&  \ce{^{72}Ga} & \phn2.05\% & \\
&  \ce{^{135}I} & \phn2.65\% &&& \ce{^{129}Te} & \phn2.25\% &&& \ce{^{112}Ag} &
\phn2.96\% &&&  \ce{^{72}Ga} & \phn2.95\% &&&  \ce{^{77}Ge} & \phn2.02\% & \\
 \bottomrule
\end{tabularx}
\begin{tabularx}{0.9\textwidth}{>{\raggedright\arraybackslash}X}
${}^\text{a}$ Symmetric fission reactions that do not create free neutrons. \\
${}^\text{b}$ The $\bar{f}_i$'s shown in these columns are averaged over all
nucleosynthesis calculations (with different initial electron fractions,
entropies, and expansion timescales) whose $Y_e$ falls within the $Y_e$ bin. \\
${}^\text{c}$ The $\bar{f}_i$'s shown in this column are averaged over the
entire
parameter space. \\
\end{tabularx}
\end{table*}

To determine the particular nuclei that are likely to power kilonovae, we
integrate the fractional heating contributions of all nuclides to find out which
nuclides contribute most to the heating. For a single nucleosynthesis
calculation, we know the total heating rate $\epsilon(t)$ as a function of time
and we can calculate the heating rate $\epsilon_i(t)$ due to nuclide $i$ as a
function of time. $\epsilon_i(t)$ is calculated as
\begin{align}
  \epsilon_i(t) = N_A\sum_{\alpha\in\mathcal{D}_i} \lambda_\alpha(t) Q_\alpha
Y_i(t), \label{eq:epsi}
\end{align}
where $\alpha$ is an index of a reaction in the reaction network and it runs
over the set $\mathcal{D}_i$, which is the set of all reactions that destroy
exactly one nuclide $i$. $N_A$ is the Avogadro constant in
$\text{baryon}\,\text{g}^{-1}$, $\lambda_\alpha(t)$ is the reaction rate of
reaction $\alpha$ in $\text{s}^{-1}$, $Q_\alpha$ is the energy released in
reaction $\alpha$ in erg, and $Y_i(t)$ is the number abundance of nuclide $i$ in
$\text{baryon}^{-1}$. Note that the total heating rate is \smash{$\epsilon(t) =
\sum_i \epsilon_i(t)$}, where $i$ runs over all nuclear species in the network.

At any given time $t$, we can now calculate the fractional heating contribution
of nuclide $i$ as \smash{$\epsilon_i(t)/\epsilon(t)$}, which is the fraction of
the total heating rate at time $t$ that is solely due to the decay of nuclide
$i$. These fractional heating contributions tell us which nuclides dominate the
heating at a given time. To quantify which nuclides dominate the heating over a
period of time, we define the \emph{integrated fractional heating
contribution} $f_i$ as
\begin{align}
f_i = \frac{1}{\ln t_1/t_0}\int_{t_0}^{t_1}
\frac{\epsilon_i(t)}{\epsilon(t)}\,d\ln t\,, \label{eq:fi}
\end{align}
where $t_0 = 0.1\,\text{days}$ and $t_1 = 100\,\text{days}$ are the beginning
and end of our heating rate fit window. We integrate in
logspace to equally weigh contributions at early and late times. Since we know
$\epsilon_i$ and $\epsilon$ only at specific time steps $t_k$, we approximate
the integral as
\begin{align}
f_i \sim \frac{1}{\ln t_1/t_0}\sum_{t_0\leq t_k \leq t_1}
\frac{\epsilon_i(t_k)}{\epsilon(t_k)}
\ln\frac{t_{k+1}}{t_k}\,.
\end{align}
If no $t_k$ is equal to $t_0$ or $t_1$, we add these two endpoints to the sum
and interpolate $\epsilon_i$ and $\epsilon$ at those points.

Note that we calculate $f_i$ for each nuclide $i$ in a single nucleosynthesis
calculation. So we should really say $f_i(Y_e,s,\tau)$, because $f_i$ will be
different for the same nuclide $i$ in different nucleosynthesis calculations
since different amounts of nuclide $i$ are be produced, depending on $Y_e$,
$s$, and $\tau$. To get an idea of which nuclides have the biggest influence on
the heating rate over a range of $Y_e$, $s$, and $\tau$, we average $f_i$ over
multiple nucleosynthesis calculations in our parameter space. We call this the
\emph{average integrated fractional heating contribution} $\bar{f}_i$ and
calculate it as
\begin{align}
  \bar{f}_i =
\frac{1}{|\mathcal{Y}|\,|\mathcal{S}|\,|\mathcal{T}|}\sum_{Y_e\in\mathcal{Y}}
\sum_ {s\in\mathcal{S}}\sum_{\tau\in\mathcal{T}} f_i(Y_e,s,\tau),
\label{eq:fibar}
\end{align}
where $\mathcal{Y}$, $\mathcal{S}$, and $\mathcal{T}$ are the sets of values of
$Y_e$, $s$, and $\tau$, respectively, that we are averaging over, and
$|\mathcal{Y}|$, $|\mathcal{S}|$, and $|\mathcal{T}|$ are the cardinalities of
those sets, i.e.\ the number of elements in the sets. Note that this method of
averaging is meaningful because we are considering the fractional heating
contribution of nuclide $i$ and not the absolute heating contribution, and
furthermore, we normalize $f_i(Y_e,s,\tau)$ in the same way for each
nucleosynthesis calculation. The final number $\bar{f}_i$ that we obtain is a
number between 0 and 1 and it tells us that nuclide $i$ is responsible for this
fraction of the total heating rate between 0.1 and $100\,\text{days}$ averaged
over a certain set of parameters $Y_e$, $s$, and $\tau$. Note that $\bar{f}_i$
is not intended to be used to estimate the absolute amount of heating due
to nuclide $i$, because the absolute amount of heating can vary greatly between
the different nucleosynthesis cases over which we averaged to obtain
$\bar{f}_i$. Rather, $\bar{f}_i$ is intended to quantify how important
different nuclides are in the makeup of the total radioactive heating rate over
a wide range of possible kilonovae. This can help inform experiments that are
measuring the $\beta$-decay properties of nuclides produced in the r-process.
To model the r-process and associated kilonovae more accurately, it would be
more beneficial to have precise measurements of the $\beta$-decay properties of
nuclides that have a larger $\bar{f}_i$ than of nuclides with smaller
$\bar{f}_i$.

\Cref{tab:dominant_nuclides} shows the 10 most dominant heating nuclides and
their average integrated fractional heating contributions $\bar{f}_i$. The
$\bar{f}_i$'s are
averaged over different high-resolution \verb\sym0\ (symmetric fission with no
free neutrons) runs in different $Y_e$ bins and over the entire range of
entropies ($1\,\kb \leq s \leq 100\,\kb$) and expansion timescales
($0.1\,\text{ms}\leq\tau\leq500\,\text{ms}$). In each $Y_e$ bin, the nuclides
are
sorted with decreasing $\bar{f}_i$. We only look at the $Y_e$-dependence of
the dominant heating nuclides because the r-process depends very strongly
on $Y_e$, while it is quite insensitive to entropy \citep[e.g.][also see
\cref{fig:abundances}]{freiburghaus:99}. Only the 10 most dominant heating
nuclides are
shown here, the full table, and the tables of the runs with different fission
reactions, are available at \web. The single most important nuclide for heating
between 0.1 and $100\,\text{days}$ is \smce{^{132}I}. It dominates over all
other nuclides by a factor of at least 3 to 10 and it especially dominates at
low initial $Y_e$. \smce{^{132}Sn} is doubly magic (50 protons and 82 neutrons)
and so it gets produced in high quantities in the r-process. Within minutes,
\smce{^{132}Sn} decays to \smce{^{132}Sb} which decays to \smce{^{132}Te}.
\smce{^{132}Te} has a half-life of $3.2\,\text{days}$ and so it decays in the
middle of our fit window where we are looking at the heating contributions. But
the decay of \smce{^{132}Te} to \smce{^{132}I} has a $Q$-value of only about
$500\,\text{keV}$, while
\smce{^{132}I} decays to the stable isotope \smce{^{132}Xe} (which is in the
middle of the second r-process peak) with a half-life of only
$2.3\,\text{hours}$ and a $Q$-value of $3.6\,\text{MeV}$. Thus we get a large
heating contribution from \smce{^{132}I}.

As is to be expected, at very low $Y_e$ (between 0 and 0.125), most of the
heating comes from nuclei that form the second ($A \sim 130$) and third ($A \sim
200$) r-process peaks.  A few very heavy nuclides ($A\sim 250$) contribute. At
higher $Y_e$ (between 0.125 and 0.25), the 10 most significantly contributing
nuclides are all in the second peak, since anything in the third peak and beyond
is more difficult to produce. The nuclides we find to be the dominant source of
heating at low initial $Y_e$ are consistent with the dominant $\beta$-decay
nuclei that \cite{metzger:10} found. They only investigated a $Y_e = 0.1$
outflow and we confirm that this result holds for a range of electron fractions
below 0.25.

At $Y_e$ between 0.25 and 0.375 there is a mix of significant contributers from
the first ($A\sim 88$) and second peaks. There are also some iron peak elements,
but most isotopes on the neutron-rich side of the iron peak have half-lives that
are either too short or too long for our fit window. Notable exceptions are
\smce{^{59}Fe}, \smce{^{66}Ni}, \smce{^{67}Cu}, and \smce{^{72}Ga}. We do indeed
see significant contributions from \smce{^{72}Ga} and \smce{^{59}Fe}. Instead of
\smce{^{66}Ni}, we see its $\beta$-decay product, \smce{^{66}Cu}, which has a
much larger $Q$-value ($2.6\,\text{MeV}$ instead of $250\,\text{keV}$) and a
half-life of $5\,\text{minutes}$. \smce{^{67}Cu} does not contribute
significantly because of its relatively low $Q$-value of $560\,\text{keV}$.
Finally, at very high $Y_e$ (between 0.375 and 0.5) there are significant
significant contributers from the proton-rich side of stability around the iron
peak.
\smce{^{57}Ni} dominates over \smce{^{56}Ni} because it has one more
neutron---thus it is a bit easier to produce in slightly neutron rich conditions
($Y_e < 0.5$)---and the $\beta^+$-decay $Q$-value of \smce{^{57}Ni} is a bit
larger than that of \smce{^{56}Ni} ($3.3\,\text{MeV}$ vs.\ $2.1\,\text{MeV}$).
Both nuclides, however, have a half-life that is right inside our fit window,
which is why both contribute significantly to the total heating rate.

The cases that produce
significant amounts of actinides also produce nuclides that undergo spontaneous
fission. In those cases, the heating due to fission becomes dominant
toward the end of the fit window (at about 100 days) but it is subdominant
throughout the rest of the fit window. The nuclides that contribute the most to
fission induced heating across the entire parameter space are \smce{^{249}Bk},
\smce{^{252}Cf}, and \smce{^{241}Pu}, which have average integrated
fractional fission
heating contributions of 33\%, 21\%, and 19\%, respectively. These numbers
are $\bar{f}_i$ defined in \cref{eq:fibar} averaged over the entire parameter
space, but the $f_i$'s of the individual nucleosynthesis calculations defined
in \cref{eq:fi} were calculated using only fission reactions in $\epsilon_i(t)$
(cf.\ \cref{eq:epsi}) and with $\epsilon(t)$ being the total heating rate due
to fission alone. In other words, averaged over all runs in the entire
parameter space and averaged in logspace over all times between 0.1 and
$100\,\text{days}$, \smce{^{249}Bk} accounts for
33\% of the entire heating due to fission, and similarly for the other
nuclides. If $\beta$-delayed
fission were included in our reaction network, it would likely significantly
alter the contribution of fission to the heating rate at low electron fraction.
For higher electron fractions, the neglect of beta-delayed fission is unlikely
to be important since very little fissible material is produced.

\section{Light curves} \label{sec:light_curves}

To test how variations in the late-time nuclear heating rate and composition
affect possible
electromagnetic transients associated with neutron star mergers, we calculate
light curves using a simplified gray radiative transport scheme in a spherically
symmetric outflow.

\subsection{Radiative transfer methods}
The ejecta is assumed to expand homologously, such that $r=vt$.  The density
structure of the outflow is then described by
\begin{equation}
\rho(t,r) = \rho_0(r/t) \left(\frac{t}{t_0} \right)^{-3}. \label{eq:rho2}
\end{equation}
\skynet\ gives a heating rate $\epsilon(t)$, which is the total amount of energy
released per unit mass and per unit time due to nuclear reactions. The majority
of this energy is carried away by neutrinos, but some fraction, say $f$, is
thermalized in the material. So $f\epsilon(t)$ is the heating rate
of the material due to nuclear reactions and decays.

For homologous outflows, the velocity can be taken as a Lagrangian coordinate.
Writing down the gray, Lagrangian radiative transport equations to first order
in $v/c$ \citep[e.g.][]{mihalas:99}, using the velocity as the Lagrangian coordinate,
and including energy release from nuclear reactions gives
\begin{align}
&\frac{d E}{d t} + \frac{2 E}{t} + \frac{1}{v^2 t} \frac{\partial}{\partial v}
\left(v^2 F \right) = \rho c \kappa \left(a T^4 - E \right),  \\
&\frac{d F}{dt}
+ \frac{1}{t} \frac{\partial}{\partial v} \left(\mathcal{F} E\right)
+ \frac{3\mathcal{F} - 1}{vt} E = - \rho c \kappa F, \\
&\frac{d u}{dt} + \frac{3P}{\rho t} = f \epsilon
+ c \kappa \left(E - a T^4 \right), \label{eq:dudt}
\end{align}
where $E$ is the radiation energy density, $t$ is the time since merger, $v$ is
the velocity measured in units of the speed of light $c$, $F$ is the
radiation flux, $\rho$ is the density given in \cref{eq:rho2}, $\kappa$ is the
opacity, $a = 4\sigma/c$ is the radiation constant where $\sigma$ is the
Stefan--Boltzmann constant, $T$ is the temperature of the fluid,  $\mathcal{F}$
is the Eddington factor (i.e.\ the ratio of the radiation pressure to the
radiation energy density), $u$ is the specific internal energy of the
fluid, $p$ is the fluid pressure, $f$ is the fraction of the heating rate
$\epsilon$ that is thermalized. The heating rate is not entirely thermalized
because a large fraction of the nuclear decay energy goes into neutrinos and gamma
rays; neutrinos are lost from the system and gamma rays are only partially thermalized.
To accurately calculate the thermalization fraction, one would need much more detailed information about
the $\beta$-decays than what is available in REACLIB and one would also have to
do $\gamma$-ray transport.
Following \cite{barnes:13}, we adopt $f = 0.3$.

The fluid is assumed to be a
non-relativistic, non-degenerate ideal gas with molecular weight $\mu$, so that
the specific internal energy is $u = 3T/(2\mu)$.  The gray transport equations
are discretized in space on a staggered grid, with $E$ and $u$ defined on zone
centers and $F$ defined on zone edges. The resulting system of ordinary
differential equations is then solved in time using a backward Euler method.
Eddington factors are obtained by solving the static Boltzmann transport
equation on a tangent ray grid at the beginning of a timestep.  This method is
similar to the one described in \cite{ensman:94}, specialized to an homologous
outflow.  The zones are chosen to be logarithmically increasing in size moving
away from the maximum radius.  This is done to ensure that the radiation
decoupling layer is resolved even at high densities.

The density structure is assumed to be described by a broken power law as argued
in \citet{chevalier:89}.  This choice was made mainly to facilitate comparison
with \cite{barnes:13}.  The power law break and density scale are fixed to give
the desired total mass and total kinetic energy of the outflow.  We use
\smash{$M =
10^{-2}\,M_\odot$} and $v = 0.1\,c$, where $c$ is the speed of light, for all
light curve models \citep[e.g.][]{hotokezaka:13,rosswog:13b,foucart:14a}.

\begin{figure*}[t!]
\includegraphics[width=\textwidth]{%
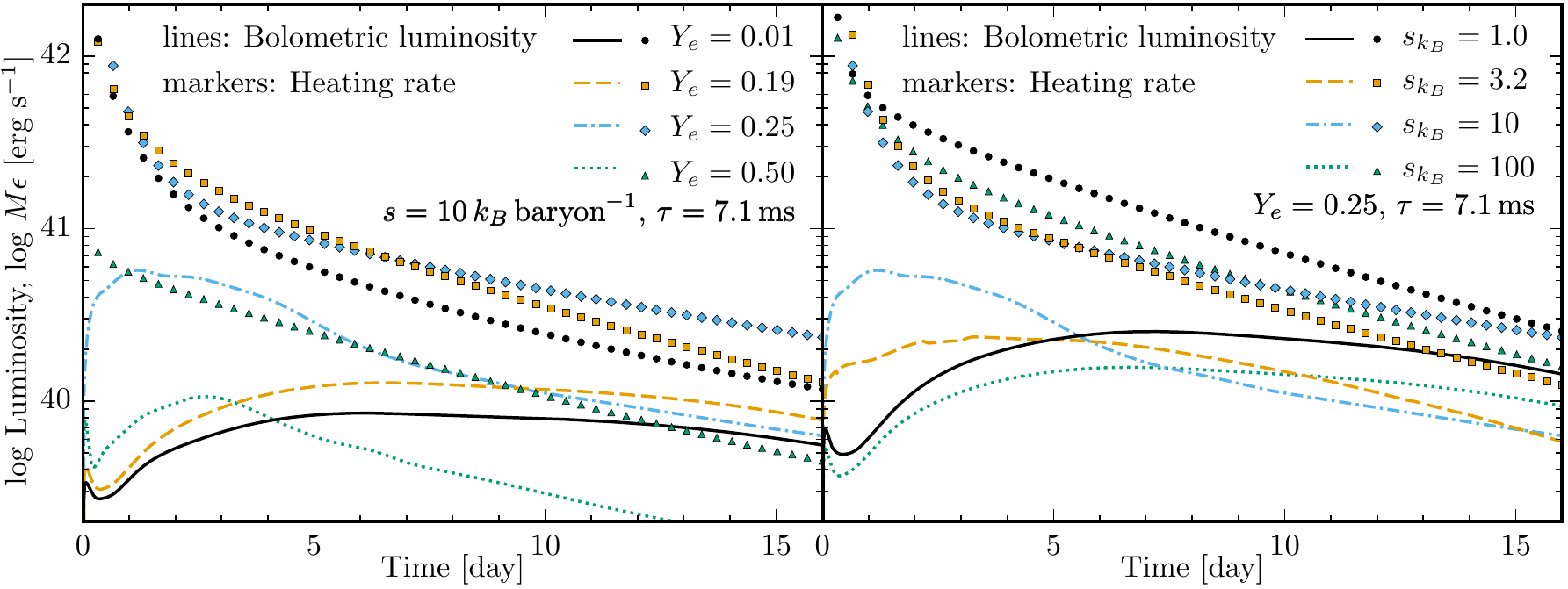}
\caption[]{The light curves and heating rates of some selected nucleosynthesis
calculations. \textbf{Left:} $Y_e = 0.01, 0.19, 0.25, 0.50$, $s = 10\,\kb$, and
$\tau = 7.1\,\text{ms}$. With $Y_e = 0.01$ and $Y_e = 0.19$ we obtain the full
r-process and so the ejecta is lanthanide-rich, which drastically increases the
opacity, resulting in a dim transient that peaks about a week after the
nucleosynthesis event. This is in contrast to the $Y_e = 0.25$ case, which has
a very similar heating rate as the low-$Y_e$ cases, but does not produce
lanthanides, and thus the transient is brighter and peaks earlier. The $Y_e =
0.50$ transient is also lanthanide-free and peaks at a few days, but because a
significant amount of stable nuclides are produced, the heating is much less,
which leads to a dim transient. \textbf{Right:} $Y_e = 0.25$, $s = 1.0, 3.2, 10,
100\,\kb$, and $\tau = 7.1\,\text{ms}$. As we saw in \cref{fig:abundances}, the
$s = 1.0\,\kb$ and $s = 100\,\kb$ cases are lanthanide-rich, while $s =
3.2\,\kb$ and $s = 10\,\kb$ are lanthanide-free, which is clearly visible in
the light curves. Even though $s = 3.2\,\kb$ and $s = 10\,\kb$ have
essentially the same heating rate, the $s = 3.2\,\kb$ case is significantly
dimmer because it has a small amount of lanthanides. The ejecta of a binary
neutron star merger is expected to have entropies between 1 and $10\,\kb$
\citep[e.g.][]{goriely:11,just:15}.}
\label{fig:light_curves}
\end{figure*}

We note that the density evolution in the transport model and the one given in
\cref{eq:rho} are both proportional to \smash{$t^{-3}$}, but they have
different scale
factors. The main point of $\rho(t)$ given in \cref{eq:rho} is to control the
timescale over which the density changes at the time of nucleosynthesis
($t\lesssim 1\,\text{s}$), but extrapolating this density to late times and
assuming that it was the uniform density of a ball of gas expanding with a
fixed velocity would lead to superluminal expansion velocities in many cases.
\Cref{eq:rho2} gives a much more reasonable estimate of the density at late
times after nucleosynthesis is over.

Calculating the exact wavelength and temperature dependent opacity of a mixture
is extremely difficult because of the large number of elements and absorption
lines involved. Especially the lanthanide and actinide element groups have very
complicated line structures and the most sophisticated line structure and
opacity calculations have only been done for a few representative nuclides
\citep[e.g.][]{kasen:13}. Such detailed opacity calculations are beyond the
scope of this work and we use a simple prescription to compute the gray opacity
$\kappa$ as a function of temperature $T$ and composition as
\begin{align}
\kappa = \kappa_\text{Fe}(T) + \sum_i \max\left[\kappa_\text{Nd}(T,X_i)
- \kappa_\text{Fe}(T),0\right],
\end{align}
where $\kappa_\text{Fe}(T)$ and $\kappa_\text{Nd}(T,X_i)$ are the iron and
neodymium opacities given in \cite{kasen:13}. The sum runs over all lanthanide
and actinide species with $X_i$ being the mass fraction of a particular
lanthanide or actinide species. We subtract the iron opacity from the neodymium
opacity because $\kappa_\text{Nd}(T,X_i)$ given in \citet{kasen:13} is actually
the opacity of a mixture containing $X_i$ neodymium and $1-X_i$ iron. Our
approximation assumes that every lanthanide and actinide contributes
the same number of lines with the same distribution in energy.  The opacity used
in the gray calculation is taken to be the Planck mean opacity, which is
appropriate when the wavelength dependent opacity is calculated in the Sobolev
approximation \citep{kasen_private:15}. At temperatures above
\smash{$10^4\,\text{K}$},
the opacities are held constant since ionization states which would have been
accessed at those temperatures were not included in the original opacity
calculation and the opacities there are artificially low
\citep{kasen_private:15}.

\subsection{Dependence of kilonova light curves on the outflow properties}

\Cref{fig:light_curves} shows the light curves and heating rates of the cases
whose final abundances are shown in \cref{fig:abundances}. In the left panel,
the
lanthanide-rich cases ($Y_e = 0.01,0.19$) are about an order of magnitude dimmer
than the lanthanide-free case ($Y_e = 0.25$) and they peak at about a week
instead of about a day. The effective temperature at peak of the
lanthanide-rich cases is also much lower ($\sim\!1600\,\text{K}$ vs.\
$\sim\!5700\,\text{K}$) than the temperature of the lanthanide-free case. The
heating rates between 0.01 and $100\,\text{days}$, however, are almost identical
for those three cases, so the significant differences in the light curves are
solely due to the amount of lanthanides present in the ejecta and their effect
on the opacity. Comparing the cases $Y_e = 0.25$ and $Y_e = 0.50$, which are
both lanthanide-free, the impact of the heating rate on the light curve can be
seen. The heating rate is lower for the $Y_e = 0.50$ case, because mostly
stable nuclei are produced, leading to less heating. The result is
that the light curve of the $Y_e = 0.50$ case peaks slightly later
($2.6\,\text{days}$ vs.\ $1.2\,\text{days}$ for $Y_e = 0.25$), is about an order
of magnitude dimmer, and redder (spectral temperature is $\sim\!3000\,\text{K}$
compared to $\sim\!5700\,\text{K}$).

In the left panel of \cref{fig:light_curves}, the light curves for $Y_e = 0.01$
and $Y_e = 0.19$ have a small peak at very early times (about
$0.04\,\text{days}$). This early peak comes from our underestimate of the
opacity at high temperatures.  There is also a small bump at early times in the
light curve of the $Y_e = 0.50$ case, which is due to the behavior of the
heating rate at early times. When determining the actual peak of the light
curve, we neglect all peaks earlier than $0.5\,\text{days}$, unless they are
more than three times brighter than all peaks after $0.5\,\text{days}$. If there
are no peaks after $0.5\,\text{days}$, we pick the brightest peak that is more
than three times brighter than the latest peak (which is also before
$0.5\,\text{days}$).

The right panel of \cref{fig:light_curves} shows selected light curves with
$Y_e = 0.25$ and various initial entropies. The cases $s = 1\,\kb$ and $s =
100\,\kb$ produce very typical lanthanide-rich light curves, whereas $s =
10\,\kb$ produces a typical lanthanide-free light curve, and $s = 3.2\,\kb$
produces a light curve that has trace amounts of lanthanides.

\begin{figure*}
\includegraphics[width=\textwidth]{%
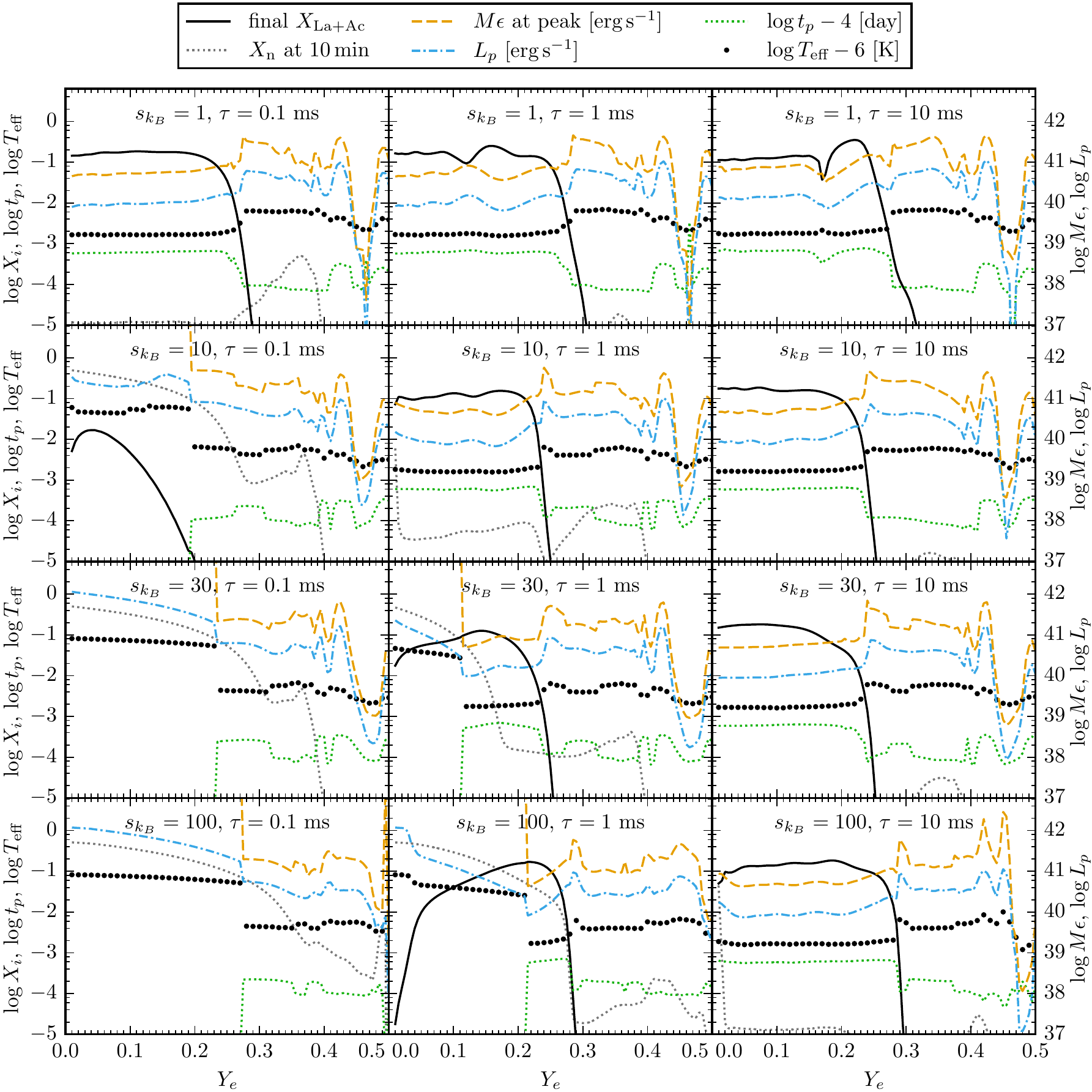}
\caption[]{The light curve results as a function of $Y_e$ for selected values
of $s$ and $\tau$. To show how lanthanides and neutron-rich freeze-out impact
the lightcurve, we again show the lanthanide and actinide
abundance $L_\text{La+Ac}$ at peak and the neutron abundance $X_n$ at
$10\,\text{minutes}$, which
were already shown in \cref{fig:hires_ye}. Additionally, we plot the heating
rate $M\epsilon$ at peak, with
\smash{$M = 10^{-2}\,M_\odot$}, the peak
luminosity $L_p$, peak time $t_p$, and the effective blackbody temperature
$T_\text{eff}$ at peak of the light curve. Note that in the
neutron-rich freeze-out cases, the heating rate $M\epsilon$ and the peak
timescale $t_p$ go off the scales, their values are \smash{$10^{44} -
10^{45}\,\text{erg}\,\text{s}^{-1}$} and $15 - 30\,\text{min}$, respectively.
As expected, $L_p$ follows
the heating rate quite closely, except in the cases where we get a neutron-rich
freeze-out. In those cases, we get a bright, very blue transient at early times.
The exact point in $Y_e$ of the transition from a neutron-powered transient to
an ordinary kilonova in this figure is somewhat arbitrary, since it depends on
the exact method
of finding the light curve peak that we choose, as explained in the text. Apart
from the
neutron-powered transients, the general trend is that we see a slightly dimmer,
redder transient at later times if the ejecta is lanthanide-rich, and a
brighter, bluer transient at earlier times if it is lanthanide-free. This is
consistent with earlier work \citep[e.g.][]{barnes:13}.}
\label{fig:hires_ye_light_curve}
\end{figure*}

In the cases where we make lanthanides at lower $Y_e$, we expect the peak
luminosity to increase and move to earlier times at higher $Y_e$ when the ejecta
transitions from lanthanide-rich to lanthanide-free, because the large
contribution to the opacity from the lanthanides suddenly goes away
\citep{kasen:13,tanaka:13}. This is shown in \cref{fig:hires_ye_light_curve}.
When lanthanides are not produced, the transient generally becomes brighter,
shorter, and bluer. We recall from \cref{fig:hires_ye} that the heating rate at
$1\,\text{day}$ tends to decrease a little when lanthanides go away. Thus the
peak luminosity $L_p$ in the lanthanide-free cases is larger not because there
is more heating in those cases, but because the peak occurs earlier (due to the
smaller opacity) and the heating rate is always larger at earlier times than at
later times.

Looking at the time $t_p$ of the light curve in \cref{fig:hires_ye_light_curve},
we see that the light curve peaks at about $6\,\text{days}$ if the ejecta is
lanthanide-rich and at about $1\,\text{day}$ if the ejecta is lanthanide-free,
which is consistent with earlier work
\citep[e.g.][]{roberts:11,barnes:13,tanaka:13}. At high $Y_e$, where we see some
oscillations in the heating rate due to specific nuclides being produced (as
explained in \sref{sec:lan_prod}), the variation in the heating rate is
reflected in the peak luminosity $L_p$ and the peak time $t_p$. More heating
results in a brighter transient at later times because the heating keeps the
ejecta hotter, and thus the opacity remains high since more excited
levels are populated, which increases the number of optically thick
lines \citep{kasen:13}. Conversely, less heating leads to
a dimmer transient at earlier times because the ejecta is cooler and thus the
opacity is lower. This variation is also reflected in the effective temperature
$T_\text{eff}$ of the transient, but to a lesser degree. In general,
lanthanide-rich transients have $T_\text{eff} \sim 1600\,\text{K}$, which peaks
at $\lambda \sim 1.8\,\upmu\text{m}$ in the infrared H and K bands.
Lanthanide-free transients have $T_\text{eff} \sim 6000\,\text{K}$ (although
this is a bit lower at very high $Y_e$ where the radioactive heating is
reduced), which peaks at $\lambda \sim 480\,\text{nm}$ in the optical B band.

In \cref{fig:hires_ye_light_curve}, we can also clearly see that neutron-rich
freeze-out produces very bright, very early, and very ultraviolet transients.
The cleanest examples are $s30\tau0.1$ and $s100\tau0.1$.
There the luminosity ranges from $2\times 10^{41}$ to
$10^{42}\,\text{erg}\,\text{s}^{-1}$, the effective temperature is about
$7\times 10^4\,\text{K}$, which peaks at $\lambda \sim 40\,\text{nm}$ (extreme
ultraviolet), and the peak occurs about an hour after the nucleosynthesis
event. These results are very similar to what \citet{metzger:14b} found,
however, they found peak effective temperatures of $\sim 10^4\,\text{K}$,
because they used higher opacities ($\kappa = 30\,\text{cm}^2\,\text{g}^{-1}$)
since their trajectories still contained a significant amount of lanthanides
and actinides \citep{metzger_private:15}. In our case, we do not find
significant amounts of lanthanides or actinides if we obtain a strong
neutron-rich freeze-out, and thus we get a lower opacity, which raises the
effective temperature \mbox{\citep{li:98}}, making such a transient even harder
to
detect because it peaks deeper in the ultraviolet. It appears that more
work is needed to consistently model these neutron-powered transients.

Note that the transition point in $Y_e$ in \cref{fig:hires_ye_light_curve}
where the light curve peaks at about
$1\,\text{hour}$ to where it peaks at a few days is somewhat arbitrary because
it depends on how we determine the peak in the light curve.  As explained above,
we arbitrarily decided to only consider peaks occurring earlier than
$0.5\,\text{days}$ if they are more than three times brighter than any later
peaks. The justification for this is that early peaks are very short and thus
hard to detect, but in the cases where we only obtain a short, bright early
peak, we do not want to pick out any later peaks that are really just the
highest points of very shallow and long plateaus.


We emphasize that the outflows used in this section were assumed to have
homogeneous compositions.  In nature, outflows from compact object mergers will
have some spread in electron fraction and therefore have inhomogeneous
compositions.  Nonetheless, our simplified models provide guidance on the
sensitivity of kilonova light curves to variations in the average electron
fraction, entropy, and dynamical timescale during r-process nucleosynthesis.

\subsection{Mass estimates of potential kilonova observations}

Since the ejecta mass is a parameter in our simplified light
curve model, we can attempt to put a lower bound on the ejecta mass necessary to
reproduce the possible kilonova observations. For the possible kilonova
associated with GRB130603B, there is one observation in the infrared, one upper
limit in the optical, and another upper limit in the infrared at late times
\citep{berger:13,tanvir:13}. For every point in our low-resolution
\texttt{sym0} parameter space we compute nine light curves with all
combinations of $v/c = 0.1, 0.2, 0.3$ and $M/M_\odot = 0.01, 0.05, 0.15$. We
then compute the observed AB magnitudes that would result from the light curve
at the rest frame time when the observations were made, taking into account
redshift and the actual filter response of the \HST{}\footnote{\url{%
http://svo2.cab.inta-csic.es/svo/theory/fps3/index.php?mode=browse&gname=HST}}.
Finally,
we interpolate the resulting observed magnitudes as a function of the ejecta
mass to find the minimum mass that reproduces the observed magnitude in the
near-infrared band (\HST{} filter WFC3/F160W, roughly J-band
in the rest frame) and produces an optical signal (\HST{} filter
WFC3/F606W, roughly B-band in rest frame) that is below the observed
upper limits.

We repeat the above procedure for light curves calculated with different
heating efficiencies $f$ (see \cref{eq:dudt}), as the exact value of $f$ is not
known but has a direct influence on the brightness of the kilonova. For $f =
0.1$, 0.3, and 0.5, we find that the minimum (over our entire parameter space)
ejecta masses necessary to match the possibly observed kilonova after GRB130603B
are 0.09, 0.03, and 0.02 solar masses, respectively. This is a reasonable
result, as we expect the minimum mass necessary to produce a kilonova of equal
brightness to decrease as the heating efficiency increases.

If we repeat the same procedure with the potentially observed kilonova after
GRB060614
(where there are detections in both the near-infrared (\HST{} filter
WFPC2/F814W, roughly R-band in rest frame) and optical (\HST{} filter
WFPC2/F606W, roughly V-band in rest frame), two infrared upper
limits at late times, and an optical upper limit at late times \citep{yang:15,
jin:15}), we find that
none of our light curves calculated with $f = 0.1$ can match the observations,
and for $f = 0.3$ and 0.5 we require a minimum mass of 0.04 and 0.05 solar
masses, respectively. We note that a larger ejecta mass is needed when a
larger heating efficiency is assumed.  Because there are observations in two
bands for GRB060614, our fits are more sensitive to the spectral temperature
found in the light curve models than in the case of GRB130603B.  Qualitatively,
the spectral temperature scales inversely with the mass and proportionally to the
heating efficiency \citep{li:98}.  Therefore, to keep a fixed spectral temperature
when increasing the heating efficiency the total mass also must be increased.
Our simple method for calculating the effective temperature is likely inadequate
for detailed confrontation with multi-band observations, so these minimum masses
are necessarily approximate.  Another issue with this method of finding the minimum
allowed mass is that the outflow does not have a homogeneous composition
\citep[e.g.][]{kasen:15,just:15,metzger:14b,wanajo:14}.  Thererfore, to acquire more
accurate estimates of the minimum ejected mass in these potential kilonova
events, more sophisticated light curve model and hydrodynamical simulations
are required.  Such an analysis
was performed in \cite{hotokezaka:13b} for GRB130603B, where they found
preferred ejecta masses between 0.02 and $0.1\,M_\odot$.

Nevertheless, the work we have presented here will be very useful to estimate
the masses and maybe even other parameters from future observations of
kilonovae. With a sophisticated radiation transport method, one can
calculate accurate light curves using our heating rates and lanthanide and
actinide abundances. A consequence of our finding that the heating rate
does not strongly depend on $Y_e$ in the lanthanide-rich regime (and not even
on $s$ and $\tau$ except at very low $Y_e$) is that one will be able to quite
accurately estimate the ejecta mass of future observed kilonovae without
precisely knowing the values of $Y_e$, $s$, and $\tau$. A caveat is, however,
that one needs to know the heating efficiency and lanthanide and
actinide opacities well.

\section{Conclusions}\label{sec:conclusion}

We have systematically performed nucleosynthesis calculations with our
new nuclear reaction network \skynet\ for a wide range of three parameters:
initial electron fraction ($0.01 \leq Y_e \leq 0.5$), initial entropy $1\,\kb
\leq s \leq 100\,\kb$, and the expansion timescale $0.1\,\text{ms} \leq \tau\leq
500\,\text{ms}$ during nuclear burning. We ran the full parameter space with
different fission reactions, but found that there were only small quantitative
and no qualitative differences between the different fission reactions. We
focused our attention on the amount of lanthanides and actinides produced and
the heating rate between 0.1 and $100\,\text{days}$ after the start of the
nucleosynthesis calculation, because kilonova transients are expected to occur
in this time frame. With a spherically symmetric, gray radiation transport scheme
we estimated the peak time, peak luminosity, and peak spectral temperature of the
kilonova light curves.

We find that the final amount of lanthanides and actinides depends most strongly
on $Y_e$ and the ejecta is lanthanide-free for $Y_e\gtrsim 0.26$. However, there
are some regions of the parameter space where the ejecta is lanthanide-free even
for very low electron fractions. Specifically, at high initial entropies and
small expansion timescales we get a neutron-rich freeze-out, which does not
produce lanthanides, but may result in a very bright, very blue transient on the
timescale of an hour. At small initial entropies and very large expansion
timescales, there is significant late-time heating, which causes the composition
to go back to \NSE{} and effectively restart the r-process at a much higher
electron fraction, which was raised by $\beta$-decays.

Since the lanthanides and actinides can increase the opacity of the material by
a factor of $\sim\!100$, we find that the peak luminosity increases by about one
order of magnitude and the light curve peak timescale goes from about a week to
about a day as the ejecta becomes lanthanide-free. This is consistent with
previous works by \citet{roberts:11,kasen:13,tanaka:13,grossman:14}. The
heating rate at
$1\,\text{day}$, however, remains largely unchanged and decreases by no more
than one order of magnitude as the ejecta becomes lanthanide-free. Thus the
increase in the kilonova luminosity is due to the decrease in the opacity when
lanthanides are no longer present, which pushes the peak to earlier times when
the heating is stronger. At very high $Y_e$ ($\gtrsim 0.4$), there are large
variations in the heating rate because single nuclides dominate the heating. At
lower $Y_e$, the heating rate at $1\,\text{day}$ is very uniform in entropy and
expansion timescale because it is dominated by an ensemble of nuclides that
average out to the same heating rate at $1\,\text{day}$ even though the exact
composition may be very different. This has already been found in
\cite{metzger:10} and we are now confirming it for a larger parameter space.

Overall, we find only weak correlation between the lanthanide production and
heating rate. Both quantities are quite strongly correlated with $Y_e$, but not
so
much with one another. The heating rate at $1\,\text{day}$ is not affected much
when the lanthanide abundance suddenly drops by many order of magnitude, but it
slowly declines at higher $Y_e$.

In \sref{sec:lan_prod}, we provided three linear inequalities involving $Y_e$,
$\ln s$, and $\ln \tau$ that can be used to determine if the ejecta with those
properties is lanthanide-rich or lanthanide-free. Those inequalities give the
correct answer in 98\% of all cases. We also provide parametric fits for the
heating rates between 0.1 and $100\,\text{days}$ for all cases at \web. The mean
fractional log difference between the actual heating rate and our fit is no more
than 1\% in 95\% of all cases. On the same website, we also provide an
integrated fractional heating contribution to give an idea of which specific
nuclides contribute the most to the radioactive heating.

Our nucleosynthesis code \skynet\ will be released as free and open-source code
soon. In the meantime, those interested can contact the authors about getting
early access to the code. Future versions of \skynet\ will also include neutrino
interactions. Much more work needs to be done to accurately model the light
curves of kilonovae and especially to calculate the line structure and hence
opacity of the lanthanide and actinide elements. We hope that our heating rate
fits will be useful to other researchers to calculate kilonova light curves that
could aid with detecting such events.

\acknowledgments

We thank Dan Kasen for helpful discussions on light curve modeling and for
graciously providing us with temperature-dependent mean opacities for various
mixtures of neodymium and iron.  We thank Christian Ott for numerous useful
discussions and for a careful reading of the manuscript.  We thank Brian
Metzger for a number of useful comments on the manuscript. And we also
thank Shri Kulkarni for discussion about computing observed magnitudes.

The calculations presented here were performed on the Caltech ``Zwicky''
compute cluster (NSF MRI award No.\ PHY-0960291), on the NSF XSEDE
network under allocation TG-PHY100033, and on NSF/NCSA Blue Waters under
allocation jr6 (NSF PRAC award No.\ ACI-1440083).
Support for LR during
this work was provided by NASA through an Einstein Postdoctoral
Fellowship grant numbered PF3-140114 awarded by the Chandra X-ray
Center, which is operated by the Smithsonian Astrophysical Observatory
for NASA under contract NAS8-03060. JL is partially supported by NSF
under award Nos.\ TCAN AST-1333520, CAREER PHY-1151197, and AST-1205732,
and by the Sherman Fairchild Foundation.

\bibliographystyle{apj}
\bibliography{apj-jour,%
bibliography/nucleosynthesis_references.bib,%
bibliography/eos_references,%
bibliography/grb_references,%
bibliography/gw_detector_references,%
bibliography/NSNS_NSBH_references,%
bibliography/methods_references,%
bibliography/nu_interactions_references,%
bibliography/cosmology_references,%
bibliography/privatebibs/roberts,%
bibliography/privatebibs/jlippuner}

\begin{thebibliography}{}
\expandafter\ifx\csname natexlab\endcsname\relax\def\natexlab#1{#1}\fi

\bibitem[{{Arcones} \& {Mart{\'{\i}}nez-Pinedo}(2011)}]{arcones:11}
{Arcones}, A., \& {Mart{\'{\i}}nez-Pinedo}, G. 2011, \prc, 83, 045809,
  \href{http://arxiv.org/abs/1008.3890}{arXiv:astro-ph.SR/1008.3890}

\bibitem[{{Arcones} {et~al.}(2010){Arcones}, {Mart{\'{\i}}nez-Pinedo},
  {Roberts}, \& {Woosley}}]{arcones:10}
{Arcones}, A., {Mart{\'{\i}}nez-Pinedo}, G., {Roberts}, L.~F., \& {Woosley},
  S.~E. 2010, \aap, 522, A25,
  \href{http://arxiv.org/abs/1002.3854}{arXiv:astro-ph.SR/1002.3854}

\bibitem[{{Argast} {et~al.}(2004){Argast}, {Samland}, {Thielemann}, \&
  {Qian}}]{argast:04}
{Argast}, D., {Samland}, M., {Thielemann}, F.-K., \& {Qian}, Y.-Z. 2004, \aap,
  416, 997,
  \href{http://arxiv.org/abs/astro-ph/0309237}{arXiv:astro-ph/0309237}

\bibitem[{{Barnes} \& {Kasen}(2013)}]{barnes:13}
{Barnes}, J., \& {Kasen}, D. 2013, \apj, 775, 18,
  \href{http://arxiv.org/abs/1303.5787}{arXiv:astro-ph.HE/1303.5787}

\bibitem[{{Bauswein} {et~al.}(2013){Bauswein}, {Goriely}, \&
  {Janka}}]{bauswein:13}
{Bauswein}, A., {Goriely}, S., \& {Janka}, H.-T. 2013, \apj, 773, 78,
  \href{http://arxiv.org/abs/1302.6530}{arXiv:astro-ph.SR/1302.6530}

\bibitem[{{Berger} {et~al.}(2013){Berger}, {Fong}, \& {Chornock}}]{berger:13}
{Berger}, E., {Fong}, W., \& {Chornock}, R. 2013, \apjl, 774, L23,
  \href{http://arxiv.org/abs/1306.3960}{arXiv:astro-ph.HE/1306.3960}

\bibitem[{Burbidge {et~al.}(1957)Burbidge, Burbidge, Fowler, \&
  Hoyle}]{burbidge:57}
Burbidge, E.~M., Burbidge, G.~R., Fowler, W.~A., \& Hoyle, F. 1957, Rev. Mod.
  Phys., 29, 547,
  \href{http://doi.org/10.1103/RevModPhys.29.547}{doi:10.1103/RevModPhys.29.547}

\bibitem[{{Chevalier} \& {Soker}(1989)}]{chevalier:89}
{Chevalier}, R.~A., \& {Soker}, N. 1989, \apj, 341, 867,
  \href{http://doi.org/10.1086/167545}{doi:10.1086/167545}

\bibitem[{Cyburt {et~al.}(2010)Cyburt, Amthor, Ferguson, Meisel, Smith, Warren,
  Heger, Hoffman, Rauscher, Sakharuk, Schatz, Thielemann, \&
  Wiescher}]{cyburt:10}
Cyburt, R.~H., Amthor, A.~M., Ferguson, R., {et~al.} 2010, \apjs, 189, 240,
  \href{http://doi.org/10.1088/0067-0049/189/1/240}{doi:10.1088/0067-0049/189/1/240},
  {REACLIB is available at \url{https://groups.nscl.msu.edu/jina/reaclib/db/}}

\bibitem[{{de Jes{\'u}s Mendoza-Temis} {et~al.}(2014){de Jes{\'u}s
  Mendoza-Temis}, {Mart{\'{\i}}nez-Pinedo}, {Langanke}, {Bauswein}, \&
  {Janka}}]{mendoza:14}
{de Jes{\'u}s Mendoza-Temis}, J., {Mart{\'{\i}}nez-Pinedo}, G., {Langanke}, K.,
  {Bauswein}, A., \& {Janka}, H.-T. 2014, ArXiv e-prints,
  \href{http://arxiv.org/abs/1409.6135}{arXiv:astro-ph.HE/1409.6135}

\bibitem[{{Duez}(2015)}]{duez_private:15}
{Duez}, M. 2015, Private communication

\bibitem[{{Eichler} {et~al.}(2014){Eichler}, {Arcones}, {Kelic}, {Korobkin},
  {Langanke}, {Marketin}, {Martinez-Pinedo}, {Panov}, {Rauscher}, {Rosswog},
  {Winteler}, {Zinner}, \& {Thielemann}}]{eichler:14}
{Eichler}, M., {Arcones}, A., {Kelic}, A., {et~al.} 2014, ArXiv e-prints,
  \href{http://arxiv.org/abs/1411.0974}{arXiv:astro-ph.HE/1411.0974}

\bibitem[{{Ensman}(1994)}]{ensman:94}
{Ensman}, L. 1994, \apj, 424, 275,
  \href{http://doi.org/10.1086/173889}{doi:10.1086/173889}

\bibitem[{{Fern{\'a}ndez} \& {Metzger}(2013)}]{fernandez:13}
{Fern{\'a}ndez}, R., \& {Metzger}, B.~D. 2013, \mnras, 435, 502,
  \href{http://arxiv.org/abs/1304.6720}{arXiv:astro-ph.HE/1304.6720}

\bibitem[{Foucart {et~al.}(2014)Foucart, Deaton, Duez, {O'Connor}, Ott, Haas,
  Kidder, Pfeiffer, Scheel, \& Szilagyi}]{foucart:14a}
Foucart, F., Deaton, M.~B., Duez, M.~D., {et~al.} 2014, \prd, 90, 024026,
  \href{http://arxiv.org/abs/1405.1121}{arXiv:astro-ph.HE/1405.1121}

\bibitem[{{Freiburghaus} {et~al.}(1999){Freiburghaus}, {Rosswog}, \&
  {Thielemann}}]{freiburghaus:99}
{Freiburghaus}, C., {Rosswog}, S., \& {Thielemann}, F.-K. 1999, \apjl, 525,
  L121, \href{http://doi.org/10.1086/312343}{doi:10.1086/312343}

\bibitem[{{Fuller} {et~al.}(1982){Fuller}, {Fowler}, \& {Newman}}]{fuller:82}
{Fuller}, G.~M., {Fowler}, W.~A., \& {Newman}, M.~J. 1982, \apjs, 48, 279,
  \href{http://doi.org/10.1086/190779}{doi:10.1086/190779}

\bibitem[{{Gehrels} {et~al.}(2009){Gehrels}, {Ramirez-Ruiz}, \&
  {Fox}}]{gehrels:09}
{Gehrels}, N., {Ramirez-Ruiz}, E., \& {Fox}, D.~B. 2009, \araa, 47, 567,
  \href{http://doi.org/10.1146/annurev.astro.46.060407.145147}{doi:10.1146/annurev.astro.46.060407.145147}

\bibitem[{{Goriely} {et~al.}(2011){Goriely}, {Bauswein}, \&
  {Janka}}]{goriely:11}
{Goriely}, S., {Bauswein}, A., \& {Janka}, H.-T. 2011, \apjl, 738, L32,
  \href{http://arxiv.org/abs/1107.0899}{arXiv:astro-ph.SR/1107.0899}

\bibitem[{{Goriely} {et~al.}(2015){Goriely}, {Bauswein}, {Just}, {Pllumbi}, \&
  {Janka}}]{goriely:15}
{Goriely}, S., {Bauswein}, A., {Just}, O., {Pllumbi}, E., \& {Janka}, H.-T.
  2015, ArXiv e-prints,
  \href{http://arxiv.org/abs/1504.04377}{arXiv:astro-ph.SR/1504.04377}

\bibitem[{{Goriely} {et~al.}(2005){Goriely}, {Demetriou}, {Janka}, {Pearson},
  \& {Samyn}}]{goriely:05}
{Goriely}, S., {Demetriou}, P., {Janka}, H.-T., {Pearson}, J.~M., \& {Samyn},
  M. 2005, Nuclear Physics A, 758, 587,
  \href{http://arxiv.org/abs/astro-ph/0410429}{arXiv:astro-ph/0410429}

\bibitem[{{Grossman} {et~al.}(2014){Grossman}, {Korobkin}, {Rosswog}, \&
  {Piran}}]{grossman:14}
{Grossman}, D., {Korobkin}, O., {Rosswog}, S., \& {Piran}, T. 2014, \mnras,
  439, 757, \href{http://arxiv.org/abs/1307.2943}{arXiv:astro-ph.HE/1307.2943}

\bibitem[{{Hoffman} {et~al.}(1997){Hoffman}, {Woosley}, \& {Qian}}]{hoffman:97}
{Hoffman}, R.~D., {Woosley}, S.~E., \& {Qian}, Y.-Z. 1997, \apj, 482, 951,
  \href{http://arxiv.org/abs/astro-ph/9611097}{arXiv:astro-ph/9611097}

\bibitem[{{Hotokezaka} {et~al.}(2013{\natexlab{a}}){Hotokezaka}, {Kiuchi},
  {Kyutoku}, {Okawa}, {Sekiguchi}, {Shibata}, \& {Taniguchi}}]{hotokezaka:13}
{Hotokezaka}, K., {Kiuchi}, K., {Kyutoku}, K., {et~al.} 2013{\natexlab{a}},
  \prd, 87, 024001,
  \href{http://arxiv.org/abs/1212.0905}{arXiv:astro-ph.HE/1212.0905}

\bibitem[{{Hotokezaka} {et~al.}(2013{\natexlab{b}}){Hotokezaka}, {Kyutoku},
  {Tanaka}, {Kiuchi}, {Sekiguchi}, {Shibata}, \& {Wanajo}}]{hotokezaka:13b}
{Hotokezaka}, K., {Kyutoku}, K., {Tanaka}, M., {et~al.} 2013{\natexlab{b}},
  \apjl, 778, L16,
  \href{http://arxiv.org/abs/1310.1623}{arXiv:astro-ph.HE/1310.1623}

\bibitem[{{Jin} {et~al.}(2015){Jin}, {Li}, {Cano}, {Covino}, {Fan}, \&
  {Wei}}]{jin:15}
{Jin}, Z.-P., {Li}, X., {Cano}, Z., {et~al.} 2015, ArXiv e-prints,
  \href{http://arxiv.org/abs/1507.07206}{arXiv:astro-ph.HE/1507.07206}

\bibitem[{{Just} {et~al.}(2015){Just}, {Bauswein}, {Pulpillo}, {Goriely}, \&
  {Janka}}]{just:15}
{Just}, O., {Bauswein}, A., {Pulpillo}, R.~A., {Goriely}, S., \& {Janka}, H.-T.
  2015, \mnras, 448, 541,
  \href{http://arxiv.org/abs/1406.2687}{arXiv:astro-ph.SR/1406.2687}

\bibitem[{{Kasen}(2015)}]{kasen_private:15}
{Kasen}, D. 2015, Private communication

\bibitem[{{Kasen} {et~al.}(2013){Kasen}, {Badnell}, \& {Barnes}}]{kasen:13}
{Kasen}, D., {Badnell}, N.~R., \& {Barnes}, J. 2013, \apj, 774, 25,
  \href{http://arxiv.org/abs/1303.5788}{arXiv:astro-ph.HE/1303.5788}

\bibitem[{{Kasen} {et~al.}(2015){Kasen}, {Fern{\'a}ndez}, \&
  {Metzger}}]{kasen:15}
{Kasen}, D., {Fern{\'a}ndez}, R., \& {Metzger}, B.~D. 2015, \mnras, 450, 1777,
  \href{http://arxiv.org/abs/1411.3726}{arXiv:astro-ph.HE/1411.3726}

\bibitem[{{Korobkin} {et~al.}(2012){Korobkin}, {Rosswog}, {Arcones}, \&
  {Winteler}}]{korobkin:12}
{Korobkin}, O., {Rosswog}, S., {Arcones}, A., \& {Winteler}, C. 2012, \mnras,
  426, 1940, \href{http://arxiv.org/abs/1206.2379}{arXiv:astro-ph.SR/1206.2379}

\bibitem[{{Kulkarni}(2005)}]{kulkarni:05}
{Kulkarni}, S.~R. 2005, ArXiv Astrophysics e-prints,
  \href{http://arxiv.org/abs/astro-ph/0510256}{arXiv:astro-ph/0510256}

\bibitem[{{Langanke} \& {Mart{\'{\i}}nez-Pinedo}(2000)}]{langanke:00}
{Langanke}, K., \& {Mart{\'{\i}}nez-Pinedo}, G. 2000, Nuclear Physics A, 673,
  481, \href{http://arxiv.org/abs/nucl-th/0001018}{arXiv:nucl-th/0001018}

\bibitem[{{Lattimer} {et~al.}(1977){Lattimer}, {Mackie}, {Ravenhall}, \&
  {Schramm}}]{lattimer:77}
{Lattimer}, J.~M., {Mackie}, F., {Ravenhall}, D.~G., \& {Schramm}, D.~N. 1977,
  \apj, 213, 225, \href{http://doi.org/10.1086/155148}{doi:10.1086/155148}

\bibitem[{{Lee} \& {Ramirez-Ruiz}(2007)}]{lee_rev:07}
{Lee}, W.~H., \& {Ramirez-Ruiz}, E. 2007, New Journal of Physics, 9, 17,
  \href{http://arxiv.org/abs/astro-ph/0701874}{arXiv:astro-ph/0701874}

\bibitem[{{Li} \& {Paczy{\'n}ski}(1998)}]{li:98}
{Li}, L.-X., \& {Paczy{\'n}ski}, B. 1998, \apjl, 507, L59,
  \href{http://arxiv.org/abs/astro-ph/9807272}{arXiv:astro-ph/9807272}

\bibitem[{Lippuner \& Roberts(2015)}]{lippuner:15b}
Lippuner, J., \& Roberts, L.~F. 2015, in preparation

\bibitem[{{Martin} {et~al.}(2015){Martin}, {Perego}, {Arcones}, {Thielemann},
  {Korobkin}, \& {Rosswog}}]{martin:15}
{Martin}, D., {Perego}, A., {Arcones}, A., {et~al.} 2015, ArXiv e-prints,
  \href{http://arxiv.org/abs/1506.05048}{arXiv:astro-ph.SR/1506.05048}

\bibitem[{{Metzger}(2015)}]{metzger_private:15}
{Metzger}, B. 2015, Private communication

\bibitem[{{Metzger} {et~al.}(2015){Metzger}, {Bauswein}, {Goriely}, \&
  {Kasen}}]{metzger:14b}
{Metzger}, B.~D., {Bauswein}, A., {Goriely}, S., \& {Kasen}, D. 2015, \mnras,
  446, 1115, \href{http://arxiv.org/abs/1409.0544}{arXiv:astro-ph.HE/1409.0544}

\bibitem[{{Metzger} \& {Fern{\'a}ndez}(2014)}]{metzger:14}
{Metzger}, B.~D., \& {Fern{\'a}ndez}, R. 2014, \mnras, 441, 3444,
  \href{http://arxiv.org/abs/1402.4803}{arXiv:astro-ph.HE/1402.4803}

\bibitem[{{Metzger} {et~al.}(2010){Metzger}, {Mart{\'{\i}}nez-Pinedo},
  {Darbha}, {Quataert}, {Arcones}, {Kasen}, {Thomas}, {Nugent}, {Panov}, \&
  {Zinner}}]{metzger:10}
{Metzger}, B.~D., {Mart{\'{\i}}nez-Pinedo}, G., {Darbha}, S., {et~al.} 2010,
  \mnras, 406, 2650,
  \href{http://arxiv.org/abs/1001.5029}{arXiv:astro-ph.HE/1001.5029}

\bibitem[{Mihalas \& Weibel-Mihalas(1999)}]{mihalas:99}
Mihalas, D., \& Weibel-Mihalas, B. 1999, {Foundations of Radiation
  Hydrodynamics} (Mineola, NY, USA: Dover Publications)

\bibitem[{M\"oller {et~al.}(2003)M\"oller, Pfeiffer, \& Kratz}]{moller:03}
M\"oller, P., Pfeiffer, B., \& Kratz, K.-L. 2003, Phys. Rev. C, 67, 055802,
  \href{http://doi.org/10.1103/PhysRevC.67.055802}{doi:10.1103/PhysRevC.67.055802}

\bibitem[{{Mumpower} {et~al.}(2012){Mumpower}, {McLaughlin}, \&
  {Surman}}]{mumpower:12}
{Mumpower}, M.~R., {McLaughlin}, G.~C., \& {Surman}, R. 2012, \prc, 85, 045801,
  \href{http://arxiv.org/abs/1109.3613}{arXiv:nucl-th/1109.3613}

\bibitem[{{Nakar}(2007)}]{nakar:07a}
{Nakar}, E. 2007, \physrep, 442, 166,
  \href{http://arxiv.org/abs/astro-ph/0701748}{arXiv:astro-ph/0701748}

\bibitem[{{Nakar} \& {Piran}(2011)}]{nakar:11}
{Nakar}, E., \& {Piran}, T. 2011, \nat, 478, 82,
  \href{http://arxiv.org/abs/1102.1020}{arXiv:astro-ph.HE/1102.1020}

\bibitem[{{Oda} {et~al.}(1994){Oda}, {Hino}, {Muto}, {Takahara}, \&
  {Sato}}]{Oda:94}
{Oda}, T., {Hino}, M., {Muto}, K., {Takahara}, M., \& {Sato}, K. 1994, Atomic
  Data and Nuclear Data Tables, 56, 231,
  \href{http://doi.org/10.1006/adnd.1994.1007}{doi:10.1006/adnd.1994.1007}

\bibitem[{{Perego} {et~al.}(2014){Perego}, {Rosswog}, {Cabez{\'o}n},
  {Korobkin}, {K{\"a}ppeli}, {Arcones}, \& {Liebend{\"o}rfer}}]{perego:14}
{Perego}, A., {Rosswog}, S., {Cabez{\'o}n}, R.~M., {et~al.} 2014, \mnras, 443,
  3134, \href{http://arxiv.org/abs/1405.6730}{arXiv:astro-ph.HE/1405.6730}

\bibitem[{{Ramirez-Ruiz} {et~al.}(2015){Ramirez-Ruiz}, {Trenti}, {MacLeod},
  {Roberts}, {Lee}, \& {Saladino-Rosas}}]{ramirez-ruiz:15}
{Ramirez-Ruiz}, E., {Trenti}, M., {MacLeod}, M., {et~al.} 2015, \apjl, 802,
  L22, \href{http://arxiv.org/abs/1410.3467}{arXiv:1410.3467}

\bibitem[{{Richers} {et~al.}(2015){Richers}, {Kasen}, {O'Connor}, {Fernandez},
  \& {Ott}}]{richers:15}
{Richers}, S., {Kasen}, D., {O'Connor}, E., {Fernandez}, R., \& {Ott}, C. 2015,
  ArXiv e-prints,
  \href{http://arxiv.org/abs/1507.03606}{arXiv:astro-ph.HE/1507.03606}

\bibitem[{{Roberts} {et~al.}(2011){Roberts}, {Kasen}, {Lee}, \&
  {Ramirez-Ruiz}}]{roberts:11}
{Roberts}, L.~F., {Kasen}, D., {Lee}, W.~H., \& {Ramirez-Ruiz}, E. 2011, \apjl,
  736, L21, \href{http://arxiv.org/abs/1104.5504}{arXiv:astro-ph.HE/1104.5504}

\bibitem[{Rosswog(2013)}]{rosswog:13b}
Rosswog, S. 2013, Phil. Trans. R. Soc. A, 371,
  \href{http://doi.org/10.1098/rsta.2012.0272}{doi:10.1098/rsta.2012.0272}

\bibitem[{{Shen} {et~al.}(2014){Shen}, {Cooke}, {Ramirez-Ruiz}, {Madau},
  {Mayer}, \& {Guedes}}]{shen:14}
{Shen}, S., {Cooke}, R., {Ramirez-Ruiz}, E., {et~al.} 2014, ArXiv e-prints,
  \href{http://arxiv.org/abs/1407.3796}{arXiv:astro-ph.GA/1407.3796}

\bibitem[{{Tanaka} \& {Hotokezaka}(2013)}]{tanaka:13}
{Tanaka}, M., \& {Hotokezaka}, K. 2013, \apj, 775, 113,
  \href{http://arxiv.org/abs/1306.3742}{arXiv:astro-ph.HE/1306.3742}

\bibitem[{{Tanvir} {et~al.}(2013){Tanvir}, {Levan}, {Fruchter}, {Hjorth},
  {Hounsell}, {Wiersema}, \& {Tunnicliffe}}]{tanvir:13}
{Tanvir}, N.~R., {Levan}, A.~J., {Fruchter}, A.~S., {et~al.} 2013, \nat, 500,
  547, \href{http://arxiv.org/abs/1306.4971}{arXiv:astro-ph.HE/1306.4971}

\bibitem[{{The LIGO Scientific Collaboration}(2015)}]{aligo}
{The LIGO Scientific Collaboration}. 2015, Class. Quantum Grav., 32, 074001,
  \href{http://arxiv.org/abs/1411.4547}{arXiv:gr-qc/1411.4547}

\bibitem[{{Timmes} \& {Swesty}(2000)}]{timmes:00}
{Timmes}, F.~X., \& {Swesty}, F.~D. 2000, \apjs, 126, 501,
  \href{http://doi.org/10.1086/313304}{doi:10.1086/313304}

\bibitem[{{van de Voort} {et~al.}(2015){van de Voort}, {Quataert}, {Hopkins},
  {Kere{\v s}}, \& {Faucher-Gigu{\`e}re}}]{vandevoort:15}
{van de Voort}, F., {Quataert}, E., {Hopkins}, P.~F., {Kere{\v s}}, D., \&
  {Faucher-Gigu{\`e}re}, C.-A. 2015, \mnras, 447, 140,
  \href{http://arxiv.org/abs/1407.7039}{arXiv:astro-ph.GA/1407.7039}

\bibitem[{Wanajo {et~al.}(2014)Wanajo, Sekiguchi, Nishimura, Kiuchi, Kyutoku,
  \& Shibata}]{wanajo:14}
Wanajo, S., Sekiguchi, Y., Nishimura, N., {et~al.} 2014, The Astrophysical
  Journal, 789, L39,
  \href{http://arxiv.org/abs/1402.7317}{arXiv:astro-ph.SR/1402.7317}

\bibitem[{{Woosley} \& {Hoffman}(1992)}]{woosley:92}
{Woosley}, S.~E., \& {Hoffman}, R.~D. 1992, \apj, 395, 202,
  \href{http://doi.org/10.1086/171644}{doi:10.1086/171644}

\bibitem[{{Yang} {et~al.}(2015){Yang}, {Jin}, {Li}, {Covino}, {Zheng},
  {Hotokezaka}, {Fan}, {Piran}, \& {Wei}}]{yang:15}
{Yang}, B., {Jin}, Z.-P., {Li}, X., {et~al.} 2015, Nature Communications, 6,
  7323, \href{http://arxiv.org/abs/1503.07761}{arXiv:astro-ph.HE/1503.07761}

\end{thebibliography}

\end{document}